\documentclass[runningheads]{llncs}
\usepackage{graphicx}
\usepackage[utf8]{inputenc}
\usepackage[T1]{fontenc}

\usepackage[portuguese]{babel}

\usepackage{hyphenat}
\usepackage{booktabs}    
\usepackage{paralist}    
\usepackage{tabularx}    
\usepackage{adjustbox}
\usepackage{array}
\usepackage{enumitem}
\usepackage{courier}
\usepackage{makecell}
\usepackage{tikz}
\usepackage{footnote}
\usepackage{appendix}
\usepackage{float}
\usepackage{listings}
\usepackage{subfigure}
\usepackage{multirow}
\usepackage{url}

\makesavenoteenv{tabular}

\def\signed #1{{\leavevmode\unskip\nobreak\hfil\penalty50\hskip2em
   \hbox{}\nobreak\hfil(#1)%
   \parfillskip=0pt \finalhyphendemerits=0 \endgraf}}
 
\newsavebox\mybox
\newenvironment{aquote}[1]
   {\savebox\mybox{#1}\begin{quotation}}
   {\signed{\usebox\mybox}\end{quotation}}
   
\def\checkmark{\tikz\fill[scale=0.2](0,.35) -- (.25,0) -- (1,.7) -- (.25,.15) -- cycle;}

\begin{document}

\mainmatter  

\title{Uma análise comparativa de \\ ferramentas de análise estática \\ para deteção de erros de memória}

\titlerunning{Uma análise comparativa de ferramentas de análise estática}

\author{Patrícia Monteiro, João Lourenço, e António Ravara}

\authorrunning{P. Monteiro \and J. M. Lourenço \and A. Ravara}

\institute{Faculdade de Ciências e Tecnologia da Universidade Nova de Lisboa (FCT-NOVA)\\
NOVA Laboratory for Computer Science and Informatics (NOVA LINCS)\\
\email{ps.monteiro@campus.fct.unl.pt} \\
\email{\{joao.lourenco, aravara\}@fct.unl.pt}}

\maketitle              

\begin{abstract}

As falhas de software estão com frequência associadas a acidentes com graves consequências económicas e/ou humanas, pelo que se torna imperioso investir na 
validação do software, nomeadamente daquele que é crítico.
Este artigo endereça a temática da qualidade do software através de uma análise comparativa da usabilidade e eficácia de quatro ferramentas de análise estática de programas em C/C++.
Este estudo permitiu compreender o grande potencial e o elevado impacto que as ferramentas de análise estática podem ter na validação e verificação de software. 
Como resultado complementar
, foram identificados novos erros em programas de código aberto e com elevada popularidade, que foram 
reportados.
\end{abstract}

\section{Introdução}

\begin{aquote}{Edsger Dijkstra, The Humble Programmer, ACM Turing Lecture 1972}
``\emph{Program testing can be a very effective way to show the presence of bugs, but is hopelessly inadequate for showing their absence}.''
\end{aquote}

A 
crescente necessidade %
de desenvolvimento de software cada vez mais complexo, no menor tempo possível e com 
baixo custo, conduz 
ao aumento da densidade de erros, 
sendo o controlo de qualidade frequentemente assegurada com base em testes e revisão manual de código.
Estas técnicas, apesar de eficazes, não permitem garantir a ausência total de erros num programa.


A presença de defeitos no software pode conduzir a problemas variados, tais como erros funcionais (o programa não cumpre os requisitos), falhas e/ou vulnerabilidades de segurança, baixa performance ou a interrupção da execução do programa. 
As falhas de software crítico são muitas vezes associadas a desastres com graves consequências económicas e/ou humanas.
São exemplos disso casos como a autodestruição do Ariane 5 (1996)\footnote{\url{https://around.com/ariane.html}} e do Mars Climate Orbiter (1999)~\cite{board_mars_1999}, o bloqueio do terminal 5 no Aeroporto de Londres-Heathrow (2008)\footnote{\url{http://www.computerweekly.com/news/2240086013/British-Airways-reveals-what-went-wrong-with-Terminal-5}} e, mais recentemente, o erro encontrado pela Amazon no Jedis (2018)\footnote{\url{https://github.com/xetorthio/jedis/issues/1747}}. 

No software escrito em C/C++ é comum a existência de erros de memória, tais como desreferências inválidas, acesso a variáveis não inicializadas e fugas e operações inválidas de libertação de memória. Este tipo de erros são de difícil deteção pois normalmente conduzem ao comportamento indefinido do programa~\cite{john_guide_2010}, isto é, tanto podem causar a sua falha imediata como permitir que este continue a funcionar de forma silenciosamente defeituosa. 

A verificação e validação de software é feita, essencialmente, através de 
três técnicas: revisão manual de código~\cite{Trisha2016}, análise %
automática (dinâmica~\cite{ball_1999} 
ou estática~\cite{wichmann_industrial_1995}) e semi-automática com 
provadores de teoremas~\cite{duffy_1991}
. 
As duas últimas são rigorosas e permitem analisar a totalidade do código, verificando 
propriedades que são definidas matematicamente e 
identificando todas as situações em que as mesmas são violadas: \emph{os erros}. 


A análise automática de software exige menos esforço por parte das equipas de desenvolvimento, fator importante no contexto atual da industria. A análise dinâmica implica a execução do programa ou de uma sua representação, enquanto que a estática analisa o programa sem executar o código fonte ou uma sua representação. 
%
A utilização precoce de técnicas de análise estática permite a identificação de erros numa fase inicial do desenvolvimento, o que reduz consideravelmente os custos
~\cite{patton_software_2006}. 
%
Como estas ferramentas utilizam frequentemente um processo de sobre-aproximação das propriedades que pretendem verificar, reportam com frequência \emph{falsos positivos}, i.e., erros que não existem.
Os falsos positivos requerem um tratamento adicional que tem custos não negligenciáveis e cuja filtragem automática pode facilmente gerar \emph{falsos negativos}, ou seja, erros reais que não são reportados. 

Interessa-nos estudar a viabilidade de usar 
ferramentas automáticas para procurar erros de memória, e o  
objetivo deste artigo é apresentar uma análise comparativa preliminar de quatro ferramentas de análise estática de código aberto: Cppcheck\footnote{\url{https://github.com/danmar/cppcheck}}~\cite{cppcheck_manual_2017}, Clang Static Analyzer\footnote{\url{http://clang.llvm.org/}}, Infer\footnote{\url{https://github.com/facebook/infer}}~\cite{calcagno_infer_2011} e Predator\footnote{\url{https://github.com/kdudka/predator}}~\cite{dudka_byte-precise_2013}. O processo e critérios de seleção destas ferramentas encontra-se descrito na Secção~\ref{subsec:tools}.
Através desta análise pretende-se identificar o tipo de padrões de erros de memória que as ferramentas conseguem detetar, identificando aquelas que são funcionalmente equivalentes, por reportarem os mesmos erros, e as que são complementares, por reportarem erros distintos.

Neste artigo apresenta-se a análise de dois programas, um de pequena e outro de média dimensão, 
estando a decorrer a análise de outros dois programas de grande dimensão.
Como seria de esperar, verificou-se que havia consideravelmente mais erros de memória reportados no software de média dimensão que no de pequena dimensão. 
Naturalmente, o número de falsos positivos também aumentou. 
No entanto, através dos resultados obtidos foi possível construir exemplos mínimos que permitiram perceber de forma clara o tipo de padrões de erros identificados pelas várias ferramentas. Os padrões de erros descobertos permitiram identificar as diferenças e falhas na análise das ferramentas. Além disso, foram descobertos novos erros nos programas escrutinados, sendo que os mesmos foram reportados nos respetivos repositórios.

\section{Metodologia}

\begin{figure}[t]
    \centering
    \includegraphics[width=\linewidth]{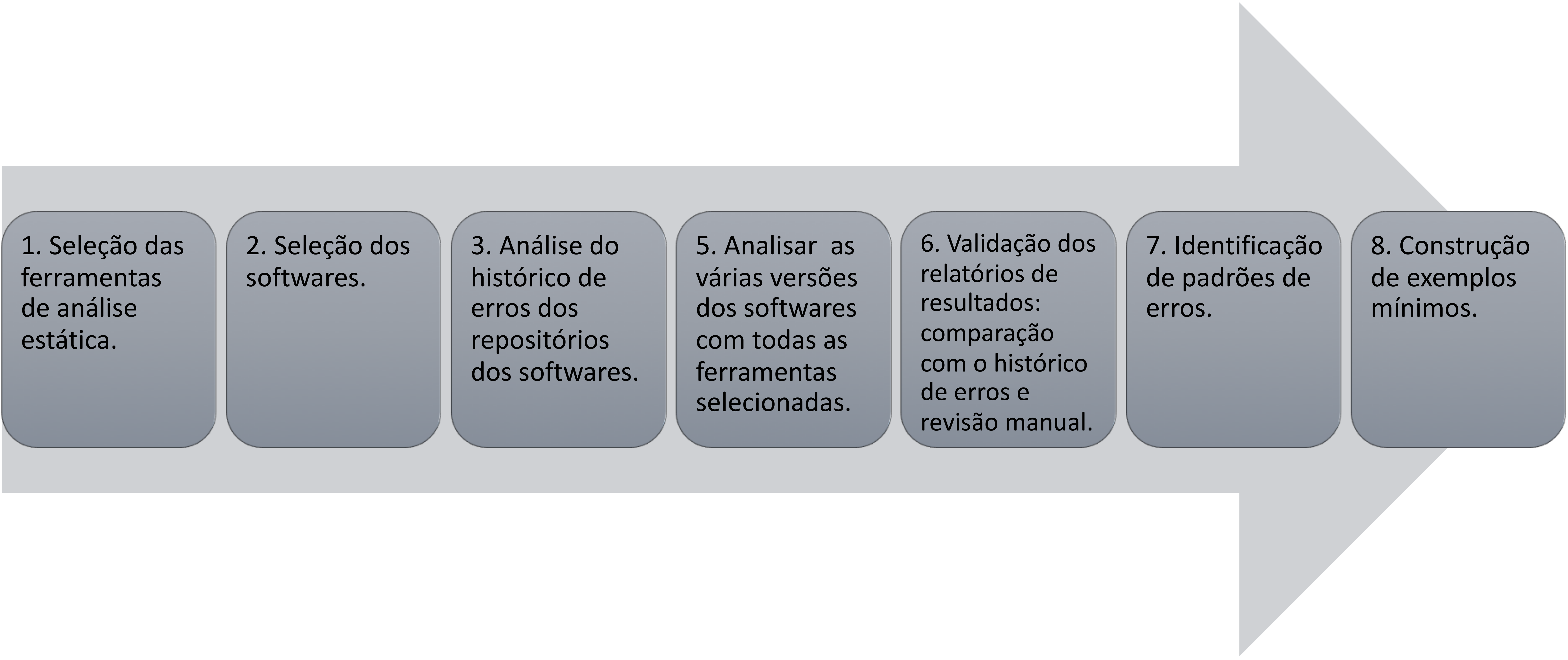}
    \caption{Etapas seguidas durante a realização da experiência.}
    \label{fig:processo}
\end{figure}

A Figura~\ref{fig:processo} apresenta o processo seguido durante este trabalho.
O processo teve início com a identificação e seleção das ferramentas de análise estática a estudar (Secção~\ref{subsec:tools}) e dos programas alvo (Secção~\ref{subsec:software}). Concluída a fase de seleção das ferramentas e dos programas, seguiu-se a análise dos mesmos. Inicialmente, foram recolhidas informações sobre o número e tipo de erros de memória já identificados nos programas, 
pesquisando os repositórios dos mesmos. Os erros reportados como questões de utilizadores (i.e., \emph{issue}) foram verificados manualmente, de maneira a verificar se são erros reais ou falsos positivos e se já estão ou não corrigidos. Por outro lado, os erros reportados como \emph{commit} foram considerados erros reais e classificados como corrigidos. As versões onde foram identificados erros foram selecionadas e analisadas por todas as ferramentas. Uma vez que as ferramentas de análise estática devolvem falsos positivos, seguiu-se a fase de validação dos seus relatórios de resultados. Esta validação, foi feita em duas etapas:
\begin{inparaenum}[i)]
    \item comparação com o histórico de erros dos repositórios; e
    \item revisão manual.
\end{inparaenum}
A comparação com o histórico de erros, para além de permitir validar resultados, também permitiu calcular a percentagem de erros identificados corretamente e detetar a existência de novos erros. 
Estes foram, posteriormente, verificados através da revisão manual do código. Por fim, 
com as informações recolhidas foi possível identificar 
padrões de erros reconhecidos pelas ferramentas e construir exemplos mínimos para cada um deles.

\subsection{Escolha de ferramentas}
\label{subsec:tools}
A escolha das ferramentas de análise estática a utilizar nesta experiência foi feita com base nos seguintes critérios: 
\begin{inparaenum}[i)]
    \item analisarem 
        programas em C/C++;
    \item 
        serem ferramentas de código aberto;
    \item 
        serem ferramentas de projetos ativos; e
    \item identificarem pelo menos dois dos seguintes erros: desreferenciação inválida, operação de libertação inválida, fuga de memória e variáveis não inicializadas. 
\end{inparaenum}

Assim, numa primeira seleção foram identificadas~32 ferramentas, que 
foram dispostas na Tabela~\ref{tab:hla:ferramentas_analise_estatica} (ver Apêndice~\ref{app:ferramentas_analise_estatica}), onde cerca de metade corresponde a ferramentas de
código aberto e outra metade a
ferramentas comerciais. 
Ficámos com uma seleção de apenas~10 ferramentas que eram simultaneamente projetos ativos e de código aberto.
%
Posteriormente, estas~10 ferramentas foram classificadas relativamente às suas funcionalidades de verificação de software (Tabela~\ref{tab:hla:funcionalidades}), tendo sido selecionadas as ferramentas que garantissem pelo menos duas das funcionalidades listadas.

\begin{table}[t]
	\caption{Funcionalidades das ferramentas de análise estática.}
	\label{tab:hla:funcionalidades}
\centering
\begin{adjustbox}{width=0.85\textwidth}
\begin{tabular}{lcccc}
	\toprule
	\multicolumn{1}{c}{\textbf{\thead[b]{Ferramenta}}} 	&  \textbf{\thead[b]{Fuga de \\ memória}}	& \textbf{\thead[b]{Desreferências \\inválidas}}	& \textbf{\thead[b]{Operações inválidas\\de libertação \\de memória}} & \textbf{\thead[b]{Variáveis \\não inicializadas}} \\
	\midrule
Cppcheck               & \checkmark  & \checkmark  & \checkmark  & \checkmark \\ 
CppLint                & x           & x           & x           & x          \\  
Clang Static An.       & \checkmark  & \checkmark  & \checkmark  & \checkmark \\ 
Cobra                  & x           & x           & x           & x          \\ 
Flawfinder             & x           & x           & x           & x          \\ 
Frama-C                & x           & \checkmark  & x           & \checkmark \\ 
Infer                  & \checkmark  & \checkmark  & \checkmark  & x          \\ 
Predator               & \checkmark  & \checkmark  & \checkmark  & x          \\ 
Uno                    & x           & \checkmark  & x           & \checkmark \\ 
VisualCodeGrepper      & x           & x           & x           & x          \\ 
    \bottomrule
\end{tabular}
\end{adjustbox}
\end{table}

Foram identificadas~6 ferramentas que cumpriam com todos os critérios referidos  (cerca de 19\texttt{\%} da lista de ferramentas inicial), que foram dispostas na Tabela~\ref{tab:hla:caracteristicas}, onde se fez a sua caracterização relativamente à facilidade de instalação, facilidade de utilização, teoria e correção. Nesta última tabela podemos observar que as ferramentas se baseiam em diferentes teorias para realizar a sua análise, sendo que a descrição de cada uma delas pode ser consultada no Apêndice~\ref{app:teorias}. Além disso, na Tabela~\ref{tab:hla:caracteristicas}, é importante ressaltar que a correção de uma ferramenta é influenciada pelo facto de esta apresentar, ou não, filtragem de falsos negativos. Relativamente à facilidade de instalação, a maioria das ferramentas tem de ser instalada manualmente. Por outro lado, todas as ferramentas estão disponíveis nas três plataformas mais comuns (Linux, macOS e Windows) excepto o Predator, que apenas funciona em Linux. Relativamente à facilidade de utilização, metade das ferramentas selecionadas não precisa de uma função \texttt{main} e apenas o Predator requere anotações adicionais no código fonte, que apenas são necessárias para imprimir 
informação e não para a execução da análise. Por fim, todas as ferramentas permitem a análise de ficheiros isolados e algumas delas (Cppcheck e Frama-C) disponibilizam interface gráfica, o que facilita a sua utilização. 
\begin{table}[t]
	\caption{Características das ferramentas de análise estática.}
	\label{tab:hla:caracteristicas}
\centering
\begin{adjustbox}{width=0.9\textwidth}
\begin{tabular}{l c c @{\hspace{1em}} c c c c c c c }
	\toprule
	\multicolumn{1}{c}{\textbf{\thead[b]{Ferramenta}}} 	& \textbf{\thead[b]{Teoria}}	& \textbf{\thead[b]{Correção}} & 
	\multicolumn{2}{c}{\textbf{\thead[b]{Instalação}}} & &
	\multicolumn{4}{c}{\textbf{\thead[b]{Utilização}}}\\
	\cmidrule[0.3pt]{4-5}\cmidrule[0.3pt]{7-10}
	&  &  &  \thead[b]{Fornece \\ executável} & \thead[b]{Multi- \\ -plataforma} & & \thead[b]{Requere\\\texttt{main}} & \thead[b]{Requere\\anotações} & \thead[b]{Análise de \\ ficheiros} & \thead[b]{Interface \\ gráfica}\\
	\midrule
Cppcheck              & AST     & x          & \checkmark & \checkmark & & x          & x          & \checkmark & \checkmark \\   
\makecell{Clang Static \\ Analyzer} & CFG     & \checkmark & \checkmark & \checkmark & & x          & x          & \checkmark & x          \\ 
Frama-C               & AST     & \checkmark & x          & \checkmark & & \checkmark & x          & \checkmark & \checkmark \\ 
Infer                 & SL, AI  & \checkmark & x          & \checkmark & & x          & x          & \checkmark & x          \\ 
Predator              & SMG     & \checkmark & x          & x          & & \checkmark & \checkmark & \checkmark & x          \\  
Uno                   & CFG     & x          & x          & \checkmark & & \checkmark & x          & \checkmark & x          \\ 
	\bottomrule
\end{tabular}
\end{adjustbox}
\end{table}

No final do processo de seleção obtiveram-se as seguintes ferramentas:
\begin{inparaenum}[]
    \item Cpp\-check~\cite{cppcheck_manual_2017},
    \item Clang Static Analyzer\footnote{\url{http://clang.llvm.org/}},
    \item Frama-C~\cite{frama_c_manual_2017},
    \item Infer~\cite{calcagno_infer_2011},
    \item Predator~\cite{dudka_byte-precise_2013} e
    \item UNO~\cite{holzmann_uno_2002}.
\end{inparaenum}
Numa primeira análise experimental, não conseguimos que as ferramentas UNO e Frama-C produzissem resultados relevantes, pelo que optámos por excluir estas ferramentas da análise comparativa final.

\subsection{Escolha de 
projetos a analisar}
\label{subsec:software}

Através de uma pesquisa de software aberto no GitHub, selecionámos~16 
projetos escritos nas linguagens C/C++, 
que foram classificados e ordenados de de acordo com os seguintes critérios:
\begin{description}
    \item[Prioridade:] 
    a nossa avaliação ponderada (1 = mais prioritário) dos demais critérios, tendo em especial consideração a
    quantidade de operações de manipulação de memória (\texttt{malloc}, \texttt{calloc}, \texttt{realloc}, \texttt{free} e utilização de ponteiros) 
    e tipo de impacto que os erros têm
    .
    \item[Popularidade:] 
    determinada pela quantidade de estrelas atribuídas pelos utilizadores ao repositório no GitHub.
    \item[Número de versões:] 
    favorecemos os programas com múltiplas versões, 
    verificando se uma ferramenta identifica os erros presentes numa determinada versão e confirma que os mesmos foram corrigidos nas versões seguintes. 
    \item[Dimensão:] 
    Assumindo que 1kB corresponde aproximadamente a 
    40 linhas no programa fonte, os programas foram divididos por dimensão 
    (Tabela~\ref{tab:hla:dimensao}). Agrupámos programas com a mesma dimensão 
    para facilitar a seleção, uma vez que nos interessava testar projetos com dimensões diferentes.
    
    \begin{table}[t]
	\caption{Classificação das dimensões dos Programas.}
	\label{tab:hla:dimensao}
\centering
\begin{adjustbox}{width=0.5\textwidth}
\begin{tabular}{l@{~~~}r@{--}l@{~~~~}r@{--}l}
	\toprule
	\multirow{2}{*}{\textbf{Classificação}} & \multicolumn{4}{c}{\textbf{Dimensão}}	\\\cmidrule{2-5}
    & \multicolumn{2}{c}{(kB)~~~~~~} 
    & \multicolumn{2}{c}{(\# linhas)~~~} \\
	\midrule
Pequeno & 50 & 400  & 2000 & 6000 \\
Médio & 400 & 1600 & 6000 & 64000 \\
Grande & 16000 & 128000 & 64000 & 512000 \\ 
	\bottomrule
\end{tabular}
\end{adjustbox}
\end{table}

    \item[Erros de memória:] número e percentagem de erros de memória reportados nos \emph{commits}.
    \item[Falha:] impacto dos erros identificados no software, 
      classificado em 3 categorias (por ordem decrescente): falha, performance e segurança. 
    Consideramos mais prioritários programas com maior percentagem de falhas. 
    \item[Questões sobre erros:] 
    submetidas pelos utilizadores e referentes a erros de memória identificados durante a instalação ou utilização do software.  
    Em alguns casos, os programas poderão não ter erros de memória identificados, mas ter uma grande quantidade de questões sobre esse tipo de erros por responder ou resolver.
    \item[Manipulação de memória:] operações de manipulação de memória 
    e quantidade de ponteiros utilizados.
\end{description}

Seguindo os critérios apresentados, os programas foram agrupados por dimensão, e 
depois ordenados por prioridade. 
Os 
com a mesma dimensão e prioridade foram ordenados 
por popularidade. 
Na Tabela~\ref{tab:hla:software} estão listados, de forma já ordenada, todos os programas analisados e respetivas características. 
Os quatro escolhidos (um de pequena dimensão, um de média dimensão e dois de grande dimensão) são os que se encontram posicionados no topo de cada um grupos
, tendo eles 
prioridade igual a 1 e elevada popularidade
:
\begin{description}
    \item [Simple Dynamic String (SDS)]\hspace{-1ex}\footnote{\url{https://github.com/antirez/sds}} biblioteca de \emph{strings} dinâmicas, projetada para aumentar as funcionalidades limitadas das \emph{strings} da biblioteca C;
    \item [Beanstalkd]\hspace{-1ex}\footnote{\url{https://github.com/kr/beanstalkd}} gestor de tarefas para aplicações distribuídas;
    \item [Tmux]\hspace{-1ex}\footnote{\url{https://github.com/tmux/tmux}} multiplexador de terminal, que permite que uma série de terminais possam ser acedidos e controlados a partir de um único terminal;
    \item [Memcached]\hspace{-1ex}\footnote{\url{https://github.com/memcached/memcached}} sistema distribuído de cache em memória, que é frequentemente utilizado para acelerar sites dinâmicos, colocando os dados dos mesmos em cache para reduzir o número de vezes que uma fonte de dados externa precisa de ser acedida. 
\end{description}

\begin{table}[t]
	\caption{Características dos programas.}
	\label{tab:hla:software}
\centering
\begin{adjustbox}{width=1\textwidth}
\begin{tabular}{l c c c c c c c c c }
	\toprule
	\multicolumn{1}{c}{\textbf{\thead[b]{Software}}} 	& \textbf{\thead[b]{Prioridade}}	& \textbf{\thead[b]{Dimensão}} & \textbf{\thead[b]{Popularidade}} & \textbf{\thead[b]{Versões}} & \textbf{\thead[b]{Erros de \\ memória}} & \textbf{\thead[b]{Erros\\causam\\falha}} & \textbf{\thead[b]{Casos (\emph{issues}) \\ sobre erros \\ de memória}} & 	\multicolumn{2}{c}{\textbf{\thead[b]{Manipulação \\ de memória}}}\\
	& &  & (\# Estrelas) & & \texttt{(\%)} total erros & \texttt{(\%)} erros mem. & \texttt{(\%)} total casos & \# Operações & \# Apontadores \\
	\midrule
\makecell{Simple Dynamic \\ String}  & 1   & Pequena     & 2215 & 2 & 0          & 0          & 23 & 121 & 71 \\
TinyVM                 & 1   & Pequena     & 1254 & 1 & 25          & 4         & 25 & 64 & 115 \\
Twemcache              & 1   & Pequena     & 811  & 8 & 0          & 0          & 21 & 257 & 205 \\
GloVe                  & 2   & Pequena     & 2204 & 2 & 18          & 0         & 28 & 118 & 64 \\
Sparkey                & 2   & Pequena     & 784  & 2 & 5          & 0          & 0 & 86 & 69 \\
\midrule[0.25pt]
Beanstalkd             & 1   & Média       & 4405 & 34 & 6           & 7        & 10 & 93 & 129 \\
Openwebrtc             & 1   & Média   & 1519 & 1 & 3          & 1          & 2 & 439 & 323 \\
Redcarpet              & 2   & Média   & 4169 & 23 & 4          & 0         & 7 & 66 & 44 \\
Http-Parser            & 2   & Média   & 3816 & 20 & 5          & 0         & 5 & 30 & 96 \\
Leveldb                & 3   & Média   & 12915 & 18 & 4          & 6        & 7 & 37 & 173 \\
\midrule[0.25pt]
Tmux                   & 1   & Grande  & 9530 & 20 & 8          & 11        & 3 & 1116 & 918 \\
Memcached              & 1   & Grande  & 7452 & 73 & 3          & 7         & 9 & 591 & 395 \\
The foundation         & 1   & Grande  & 2810 & 670 & 3          & 6        & 7 & 682 & 1342 \\
Timescaledb            & 1   & Grande  & 4259 & 25 & 3          & 1         & 3 & 59 & 214 \\
Lwan                   & 2   & Grande  & 4131 & 1 & 15          & 8         & 21 & 462 & 700 \\
Ziparchive             & 2   & Grande  & 3368 & 37 & 1          & 2         & 3 & 67 & 284 \\
	\bottomrule
\end{tabular}
\end{adjustbox}
\end{table}

\section{Experimentação e análise}

\subsection{Relatório de resultados}
\label{subsec:reports}

Neste artigo apresentamos os resultados referentes a dois programas.
O SDS é um software de pequena dimensão com apenas~2 versões no seu repositório. 
Os respetivos \emph{commits} não reportam erros de memória,
mas existem vários erros deste tipo identificados por utilizadores (i.e., \emph{issue}), sendo que alguns desses erros ainda se encontram por corrigir e/ou validar. Portanto, todos os erros reportados no repositório e devolvidos pelas ferramentas foram verificados através de revisão manual de código.
O Beanstalkd é um software de dimensão média e possui~34 versões no repositório. 
As ferramentas foram executadas em~12 versões deste software onde haviam sido identificados erros de memória, permitindo assim comparar os resultados destas com o histórico de erros disponível. 

Os erros identificados no SDS e no Beanstalkd, quer pelos programadores e utilizadores, quer pelas ferramentas, encontram-se listados nas Tabelas~\ref{tab:hla:erros-sds} e~\ref{tab:hla:erros-beanstalkd} (Apêndice~\ref{app:relatorio_erros}). No total foram analisados~30 erros no SDS, dos quais~24 foram identificados pelas ferramentas, e~188 erros no Beanstalkd, dos quais~175 foram identificados pelas ferramentas. De notar que, na Tabela~\ref{tab:hla:erros-beanstalkd} foram omitidos os~135 falsos positivos identificados pelo Predator
. Na secção seguinte é feita uma descrição da análise dos relatórios obtidos.

\subsection{Análise dos resultados}
Os resultados obtidos pela execução das ferramentas de análise estática foram comparados com o histórico de erros dos repositório e os novos erros foram sujeitos a revisão manual de código. Assim, a partir da análise dos resultados foi possível calcular a percentagem de falsos positivos (FP), falsos negativos (FN), verdadeiros positivos (TP) e verdadeiros negativos (TN) identificados por cada uma das ferramentas.
Para uma determinada ferramenta a classificação dos erros em cada uma das categorias segue os seguintes critérios:

\medskip
\noindent
\begin{tabular}{r@{~—~}l}
    \textbf{Falso positivo} & erro não real mas reportado pela ferramenta \\
    \textbf{Falso negativo} & erro real mas não reportado pela ferramenta \\
    \textbf{Verdadeiro positivo} & erro real e reportado pela ferramenta \\
    \textbf{Verdadeiro negativo} & erro não real e não reportado pela ferramenta \\
\end{tabular}
\medskip

Um erro foi considerado real quando está reportado no histórico de erros do repositório ou %
quando é identificado por uma das ferramentas e validado 
com revisão manual de código. 

\subsubsection{Simple Dynamic String (SDS).}
\label{subsubsec:sds}
\begin{figure}[t]
    \centering
    \subfigure[SDS.]{\label{fig:per_error_sds}\includegraphics[width=0.49\linewidth]{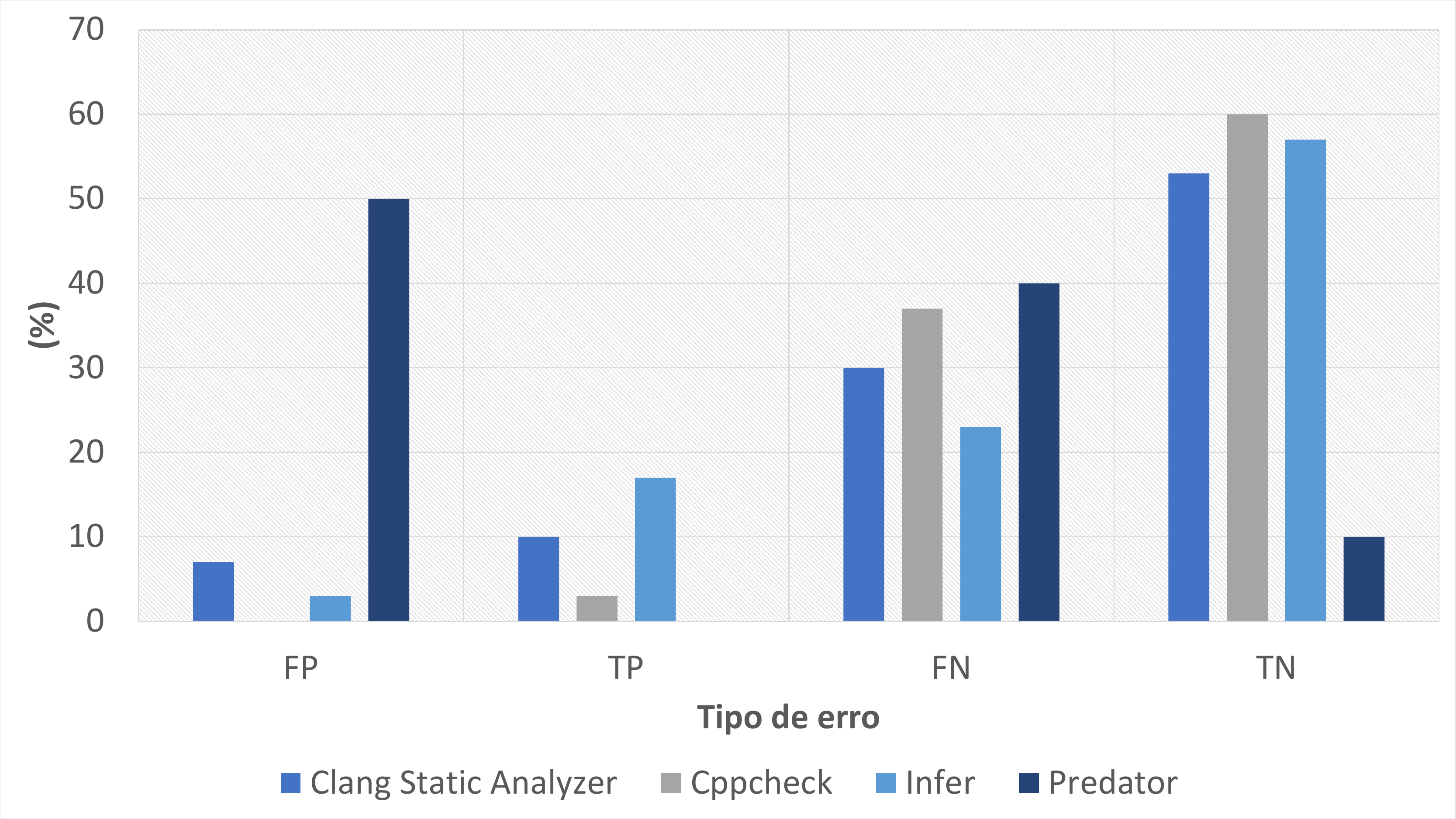}}%
    \hfill
    \subfigure[Beanstalkd.]{\label{fig:per_error_beanstalkd}\includegraphics[width=0.49\linewidth]{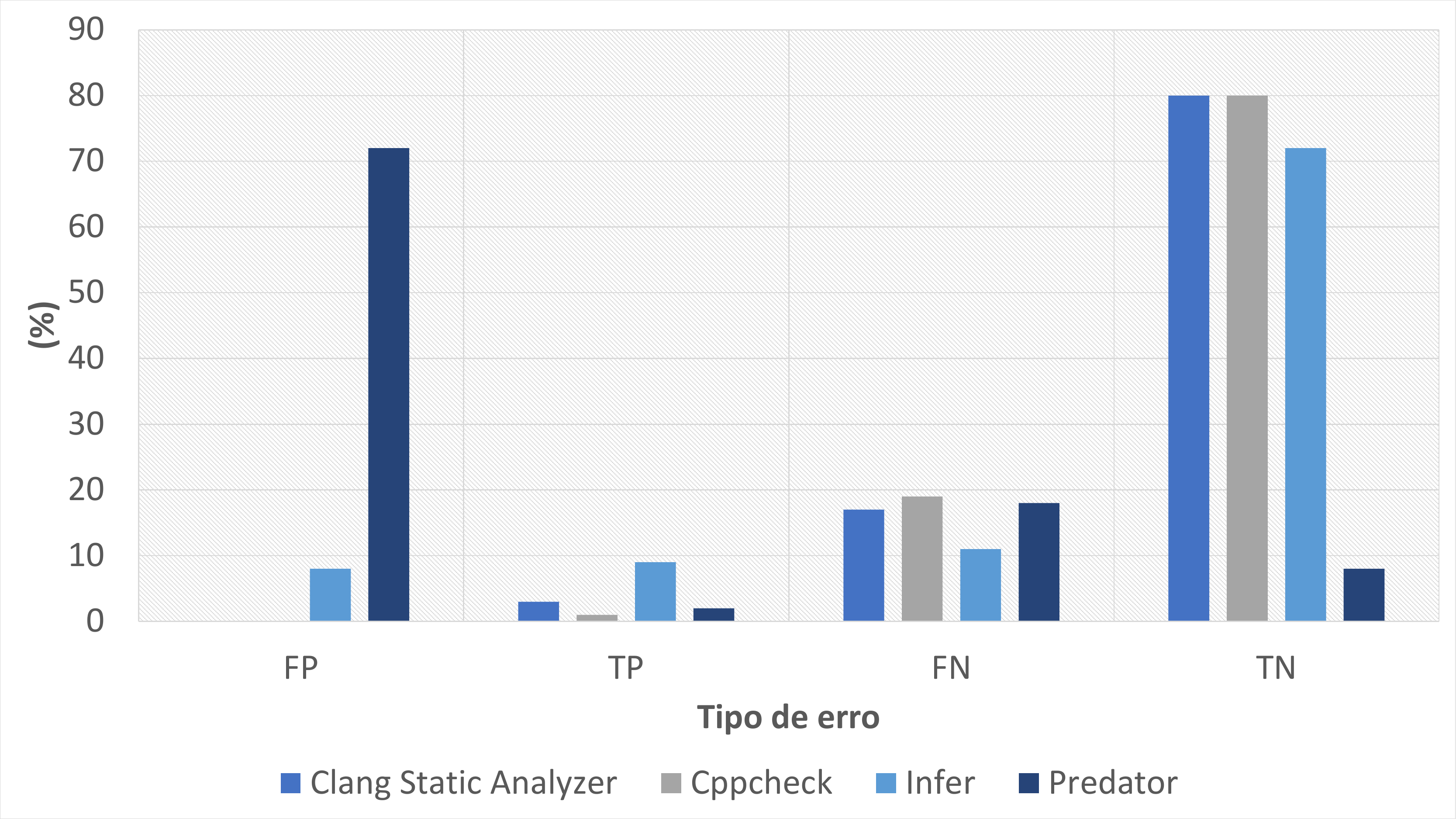}}
    \caption{Percentagem de cada tipo de erros identificado nos programas.}
    \label{fig:my_label}
\end{figure}

Na 
versão 1 
foram reportados~7 erros no repositório
, 
sendo um desses 
um falso positivo identificado por um utilizador.
Na 
versão 2 
foram reportados~2 erros, 
mas um 
já existia na versão~1. Por outro lado, as ferramentas encontram um total de~24 erros diferentes, sendo que~21 desses não foram reportados. No entanto, apenas~6 dos~21 erros poderão ser classificados como reais. Assim, no SDS foram identificados~26 erros, dos quais~13 são reais. Os erros reais 
que encontrámos na última versão 
e que reportámos 
são
:
\begin{itemize}
    \item \emph{sds.c:1123: error: null dereference}, identificado pelo Infer\footnote{\url{https://github.com/antirez/sds/issues/99}};
    \item \emph{sds.c:1240: error: dead store}, identificado pelo Clang 
    e pelo Infer\footnote{\url{https://github.com/antirez/sds/issues/100}}.
\end{itemize}
Na Figura~\ref{fig:per_error_sds} está representado um gráfico com as percentagens dos vários tipos de erros identificados pelas ferramentas no SDS. 
Este software retorna, intencionalmente, apontadores para o meio de blocos de memória alocados com \texttt{malloc}. Portanto, as operações de desreferência ou libertação de memória aplicadas a ponteiros nestas situações são reportadas pelas ferramentas Clang Static Analyzer e Predator como erros, que nós classificámos como falsos positivos.

Verificámos que o Clang Static Analyzer apresentou uma percentagem relativamente baixa de falsos positivos, 
mas verificou-se também~30\texttt{\%} de falsos negativos. Esta percentagem deve-se ao facto da Clang 
não identificar um padrão de erros relevante, nomeadamente a verificação dos resultados de operações de alocação de memória (i.e., \texttt{malloc}, \texttt{calloc} e \texttt{realloc}).

Por princípio, a ferramenta Predator não permite chamadas de funções externas, a fim de excluir qualquer efeito colateral que possa potencialmente quebrar a segurança da memória. As únicas funções externas permitidas são aquelas que o Predator reconhece como funções integradas e as modela apropriadamente, provando a segurança da memória (\texttt{malloc}, \texttt{free}, e algumas funções da biblioteca do C tais como \texttt{memset}, \texttt{memcpy} e \texttt{memmove}~\cite{dudka_byte-precise_2013}). 
Por este motivo, o Predator não foi capaz de identificar nenhum erro real neste software, sendo a ferramenta que devolveu a maior percentagem de falsos positivos e falsos negativos.

Apesar de falhar na identificação de alguns erros, o Infer é a ferramenta que devolve uma maior percentagem de verdadeiros positivos e a menor percentagem de falsos negativos neste software. 
Como a análise realizada pelo Infer consiste numa execução simbólica do código, mantendo uma \emph{heap} simbólica
, quando a ferramenta não consegue provar a segurança da memória, pode reportar um erro, se encontrar uma desreferência nula ou uma fuga de memória, ou pode perceber que se encontra num estado inconsistente. Em ambos os casos, a análise é interrompida, porque a tentativa de prova não faz sentido.
Outra razão para que o Infer não consiga 
reportar erros que poderia identificar é a existência de um tempo limite para execução da análise,
 que é por vezes atingido antes da análise chegar ao 
fim~\cite{calcagno_infer_2011}, e que parece não ser possível de parametrizar.

O Cppcheck é a ferramenta que devolve a menor quantidade de erros. Por uma opção de engenharia, esta ferramenta realiza filtragem dos erros para reduzir o número de falsos positivos reportado, mas nesse processo acabar por eliminar erros reais gerando falsos negativos.

\subsubsection{Beanstalkd.}
\label{subsubsec:beanstalkd}

Este software tem uma dimensão e complexidade maior que o SDS. 
Como o Predator não está preparado para analisar programas muito complexos~\cite{dudka_byte-precise_2013}, a análise do Beanstalkd revelou-se pouco conclusiva. 
Numa tentativa de extrair algum tipo de resultado relevante, o Predator foi utilizado para testar cada um dos ficheiros do software individualmente
, tendo sido obtidos~135 falsos positivos (cerca de 72\texttt{\%} do total de erros reportados) 
que se devem às chamadas de funções externas que são ignoradas. 

O Infer foi mais uma vez a ferramenta que devolveu uma maior percentagem de verdadeiros positivos. No entanto, foi também a ferramenta com a segunda maior percentagem de falsos positivos.
%
O Cppcheck não devolveu falsos positivos, no entanto, revelou-se pouco eficaz devolvendo a menor percentagem de verdadeiros positivos e a maior percentagem de falsos e verdadeiros negativos. Estes resultados devem-se ao facto desta ferramenta fazer filtragem de falsos positivos. 
Por fim, o Clang Static Analyzer foi a ferramenta que devolveu uma maior quantidade de verdadeiros negativos (80\texttt{\%}), juntamente com o Cppcheck. 
Além disso, esta ferramenta identificou a segunda maior percentagem de verdadeiros positivos no Beanstalkd. 

Os novos erros identificados e classificados como verdadeiros positivos no repositório do Beanstalkd foram os seguintes:
\begin{itemize}
    \item \emph{prot.c:501: error: null dereference}, identificado pelo Infer\footnote{\url{https://github.com/kr/beanstalkd/issues/384}};
    \item \emph{testheap.c:222: error: memory leak}, identificado pelo Cppcheck e 
    Predator\footnote{\url{https://github.com/kr/beanstalkd/issues/382}}.
\end{itemize}
Estes erros, tal como aconteceu para o software SDS, foram reportados e aguardam \emph{feedback} por parte dos programadores responsáveis pelo repositório.



\section{Conclusões}
Como se pode verificar nas Tabelas~\ref{tab:hla:erros-sds} e
\ref{tab:hla:erros-beanstalkd} do Apêndice~\ref{app:sintese_erros}, é útil
usar as 4 ferramentas, pois obtém-se resultados complementares: cada
ferramenta identificou erros que nenhuma das outras identificou. Este
facto valida a seleção feita.

É também importante salientar que as utilização destas ferramentas não
pesa significativamente no desenvolvimento de software, pois os tempos
que cada uma leva a analisar os softwares escolhidos são muito
reduzidos, como se pode ver nas Tabelas~\ref{tab:hla:tempos_execucao_sds}
e \ref{tab:hla:tempos_execucao_beanstalkd} do Apêndice~\ref{app:tempos_execucao}:
vão de alguns segundos, no caso do SDS, a no máximo pouco mais de um minuto, no caso do Beanstalkd.

Note-se que vários dos erros encontrados pelas ferramentas estiveram
presentes nos softwares por longos períodos de tempo,
como se vê nas Tabelas~\ref{tab:hla:tempos_permanencia_sds} e \ref{tab:hla:tempos_permanencia_beanstalkd}
do Apêndice~\ref{app:tempos_permanencia}:
o SDS teve erros que foram corrigidos só passados dois anos e tem erros que presentes há cerca de quatro anos;
o Beanstalkd teve erros que foram corrigidos só passados quatro anos e tem erros presentes há quase dez anos.

A deteção destes erros
pelas ferramentas foi no entanto muito rápida (como se referiu acima) e a verificação de que se tratavam
de erros reais 
levou poucas horas. 
É então muito vantajosa a utilização destas ferramentas de análise
estática no processo de desenvolvimento de software para evitar erros
de memória.

A partir desta experiência foi possível identificar padrões de erros e construir exemplos mínimos capazes de reproduzir os resultados observados nos programas analisados.
Na Tabela~\ref{tab:hla:padroes} (Apêndice~\ref{app:patterns}) estão representados os padrões identificados para cada tipo de erro. Os exemplos mínimos e respetivos relatórios de resultados podem ser consultados no Apêndice~\ref{app:exemplos_minimos}.

A ferramenta Infer é a única que identifica a possibilidade da ocorrência de uma desreferência nula quando não é feita a verificação dos resultados das operações de alocação de memória. Se as funções \texttt{malloc}, \texttt{calloc} e \texttt{realloc} falharem, estas retornam o valor \texttt{NULL} e, portanto, a desreferência dessa variável poderá 
gerar um erro. 
O Infer é também a única ferramenta que identifica que o valor de um endereço não está a ser utilizado (i.e., \emph{dead store}), caso esse valor seja 0 ou \emph{NULL}. 
A ferramenta Clang Static Analyzer 
não considera esta situação um erro, uma vez que considera que o valor 0 ou \emph{NULL} ao ser atribuído à variável na sua inicialização não é um valor não utilizado.
O Infer (até à versão 0.13.1) não conseguia identificar qualquer tipo de erro de memória relacionado com a utilização de uma variável alocada usando o padrão \emph{sizeof(*ptr)}, e.g., \texttt{ptr = malloc(sizeof(*ptr))}.
Na versão mais recente do Infer (versão 0.14.0, lançada no dia 1 de Maio de 2018) este defeito já foi corrigido. No Apêndice~\ref{app:exemplos_minimos} deste artigo encontra-se um exemplo mínimo da situação descrita. 

Apesar de o Infer ser a única ferramenta que reporta a não verificação das chamadas a funções de alocação de memória, observou-se que o também o Cppcheck, em algumas situações, reporta este tipo erros. No entanto, os erros reportados pelo Cppcheck e Infer são diferentes. O Cppcheck identifica uma possível fuga de memória, enquanto que o Infer identifica uma desreferência nula. Isto acontece porque é feita uma atribuição sem que seja verificado o resultado da chamada à função \texttt{realloc}, ficando assim a memória anteriormente atribuída inacessível no caso de esta operação falhar.

Este tipo de ferramentas está já a ser incluído no processo de
desenvolvimento de alguns programas de grande dimensão e relevância,
como 
o 
LibreOffice\footnote{\url{https://github.com/LibreOffice/core}}. Atualmente,
este software 
usa duas ferramentas, o Cppcheck
 e o Coverity~\cite{llaguno_2017}. 
Segundo o relatório disponibilizado
, até ao momento foram analisadas mais de~6 milhões de linhas de
código do projeto LibreOffice, onde foram identificados mais de~25 mil
erros, dos quais foram
corrigidos~99\texttt{\%}~\cite{coverity_libreoffice}. No caso do
Cppcheck, segundo o último relatório disponível
 (de 27 de Janeiro de~2018), foram identificados através desta ferramenta cerca de~6 mil erros no repositório do LibreOffice~\cite{cppcheck_libreoffice}.

O trabalho futuro passa pela análise dos programas de grande dimensão já selecionados (Tmux e Memcached). Os dados recolhidos dessa análise serão depois utilizados para identificar novos padrões de erros e construir exemplos mínimos, tal como foi feito para o SDS e para o Beanstalkd. Desta forma espera-se obter novos dados para o desenvolvimento do estudo comparativo das ferramentas.

\subsubsection*{Agradecimentos.}
{\small\itshape Este trabalho foi suportado pela FCT-NOVA e parcialmente financiado pela FCT-MCTES e pelo programa POCI-COMPETE2020 nos projetos UID/CEC/04516/2013 e PTDC/CCI-COM/32456/2017.}

\bibliographystyle{alpha}
\bibliography{mybibliography}

\newpage
\appendix
\section{Escolha das ferramentas de análise estática}
\label{app:ferramentas_analise_estatica}

\begin{savenotes}
\begin{table}[h]
	\caption{Ferramentas de análise estática.}
	\label{tab:hla:ferramentas_analise_estatica}
\centering
\begin{center}  
\begin{tabular}{lcc}
	\toprule
	\multicolumn{1}{c}{\textbf{Ferramenta}} 	& \textbf{Código aberto}	& \textbf{Ativo}\\
	\midrule
AdLint\footnote{http://adlint.sourceforge.net/}                       & \checkmark  & x          \\ 
Astrée~\cite{delmas_astree_2007}                        & x           & \checkmark \\ 
Axivion Bauhaus Suite~\cite{raza_bauhaus_2006}         & x           & \checkmark \\ 
BLAST~\cite{shved_blast_2012}                         & \checkmark  & x          \\ 
Cppcheck~\cite{cppcheck_manual_2017}                      & \checkmark  & \checkmark \\
Cppdepend\footnote{https://www.cppdepend.com/}                    & x           & \checkmark \\
Cpplint\footnote{https://github.com/google/styleguide/tree/gh-pages/cpplint}                    & \checkmark  & \checkmark \\
Clang Static Analyzer\footnote{http://clang.llvm.org/}         & \checkmark  & \checkmark \\ 
Cobra~\cite{holzmann_cobra_2017}                         & \checkmark  & \checkmark \\
Codacy\footnote{https://www.codacy.com/}                       & x           & \checkmark \\ 
CodeSonar\footnote{https://www.grammatech.com/products/codesonar}                   & x           & \checkmark \\ 
Coverity~\cite{llaguno_2017}                      & x           & \checkmark \\ 
ECLAIR~\cite{bagnara_eclair_2013}                        & x           & \checkmark \\ 
Flawfinder~\cite{report_flawfinder_2017}                    & \checkmark  & \checkmark \\ 
Fluctuat~\cite{goubault_fluctuat_2008}                      & x  & \checkmark \\ 
Frama-C~\cite{frama_c_manual_2017}                       & \checkmark  & \checkmark \\ 
Goanna~\cite{fehnker_goannastatic_2007}                        & x           & \checkmark \\ 
Infer~\cite{calcagno_infer_2011}                         & \checkmark  & \checkmark \\ 
Klocwork Static Code Analysis~\cite{bolduc_klocwork_2016} & x           & \checkmark \\ 
LCLint~\cite{evans_lclint_1994}                           & \checkmark & x \\ 
LDRA Testbed~\cite{hennell_testbed_1978}                  & x           & \checkmark \\ 
Parasoft C/C++ test\footnote{https://www.parasoft.com/products/ctest}           & x           & \checkmark \\ 
PC-Lint~\cite{gimpel_pclint_2014}                       & x           & \checkmark \\ 
Polyspace~\cite{wissing_polyspace_2007}                     & x           & \checkmark \\ 
Predator~\cite{dudka_byte-precise_2013}                      & \checkmark  & \checkmark \\ 
PVS-Studio\footnote{https://www.viva64.com/en/pvs-studio/}                    & x           & \checkmark \\    
PRQA QAC ~\cite{qa-c_report_2017}            & x           & \checkmark \\ 
Saturno~\cite{xie_saturn_2005}                       & \checkmark  & x          \\
SLAM project~\cite{ball_slam_2004}                  & x           & x          \\ 
Splint~\cite{evans_splint_2002}                        & \checkmark  & x\\ 
Uno~\cite{holzmann_uno_2002}                           & \checkmark  & \checkmark \\ 
VisualCodeGrepper\footnote{https://github.com/nccgroup/VCG}           & \checkmark  & \checkmark \\
	\bottomrule
\end{tabular}
\end{center}
\end{table}
\end{savenotes}

\section{Teorias}
\label{app:teorias}

\subsection{Interpretação Abstrata (AI):} Abordagem que serve para modelar o efeito que cada declaração tem sobre o estado de uma máquina abstrata. Isto é, dada uma linguagem de programação, a interpretação abstrata consiste em atribuir várias semânticas ligadas por relações de abstração. Uma semântica é uma caracterização matemática do possível comportamento do programa. Portanto, o programa é executado com base nas propriedades matemáticas de cada declaração. Nem todas as propriedades verdadeiras do sistema original são verdadeiras no sistema abstrato, portanto, o sistema abstrato é mais simples de analisar devido à sua incompletude. Se esta abordagem for aplicada corretamente, todas as propriedades verdadeiras do sistema abstrato podem ser mapeadas para uma propriedade verdadeira do sistema original~\cite{cousot_ai_1977}.
\subsection{Árvore de Sintaxe Abstrata (AST):} Uma representação abstrata em árvore do código-fonte de um programa, na qual as folhas representam as constantes e as variáveis e os nós internos representam os operadores ou declarações~\cite{neamtiu_ast_2005}.
\subsection{Grafo de Controlo de Fluxo (CFG):} Consiste num grafo orientado, no qual são representados todos os caminhos lógicos que podem ser percorridos durante a execução de um programa~\cite{allen_cfg_1970}.
\subsection{Lógica de Separação (SL):} Uma lógica matemática, que consiste numa extensão da lógica de Hoare~\cite{hoare_1969} para o raciocínio sobre programas que acedem e alteram dados mantidos em estruturas de dados dinâmicas. Esta lógica permite a escalabilidade ao partir o raciocínio em várias partes, correspondentes às operações locais na memória e, seguidamente, juntar as partes do raciocínio novamente~\cite{reynolds_separation_2002}.
\subsection{Grafo de Memória Abstrata (SMG):} Grafo direcionado que representa simbolicamente a memória e que é composto por objetos (espaço alocado), valores (endereços) e arestas que ligam os vários elementos do grafo. Assim, cada bloco de memória utilizado pelo programa dá origem a um SMG~\cite{dudka_byte-precise_2013}.
\section{Relatórios de erros}
\label{app:relatorio_erros}

\renewcommand{\arraystretch}{1}

\begin{table}[H]
	\caption{Resultados da análise do software Beanstalkd.}
	\label{tab:hla:beanstalkd}
\centering
\begin{adjustbox}{width=1\textwidth}
\begin{tabular}{l c c c c c c}
	\toprule
	\textbf{Erro}   & 	\multicolumn{2}{c}{\textbf{Versão}} & \textbf{Reportado} & \textbf{\thead{Falso \\ positivo}} & \textbf{Ferramenta} \\
	& Identificado & Corrigido & & & \\
	\midrule
beanstalkd.c:667: error: memory leak & 0.3 & 0.4 & \checkmark & x & x \\
net.c:28: error: dead store & 0.3 & 1.4 & x & x & \thead{Clang Static An. \\ Infer} \\
beanstalkd.c:41: error: resource leak & 0.3 & 1.5 & x & x & Infer \\
reserve.c:51: error: null dereference & 0.3 & 1.5 & x & x & Infer \\ 
prot.c:140: error: null dereference & 0.3 & 1.5 & x & x & Infer \\ 
beanstalkd.c:395: error: null dereference & 0.3 & 0.5 & x & x & Infer \\ 
beanstalkd.c:737: error: memory leak & 0.3 & 0.5 & x & x & Infer \\
beanstalkd.c:205: error: memory leak & 0.4 & 0.5 & \checkmark & x & x \\
tube.c:60: error: memory leak & 0.8 & 1.2 & \checkmark & x & x \\
ms.c:56: error: memory leak & 0.8 & x & x & \checkmark & Infer \\
prot.c:320: error: null dereference & 0.8 & 1.5 & x & x & Infer \\
prot.c:374: error: null dereference & 0.8 & 1.5 & x & x & Infer \\ 
prot.c:946: error: dead store & 0.8 & x & x & \checkmark & Infer \\
prot.c:1134: error: dead store & 0.8 & x & x & \checkmark & Infer \\
prot.c:1170: error: dead store & 0.8 & x & x & \checkmark & Infer \\ 
prot.c:1230: error: dead store & 0.8 & x & x & \checkmark & Infer \\
prot.c:1388: error: dangling pointer & 1.2 & 1.3 & \checkmark & x & x \\
job.c:170: error: memory leak & 1.2 & 1.4 & \checkmark & x & x \\ 
prot.c:1493: error: memory leak & 1.2 & 1.4 & \checkmark & x & x \\
job.c:48: error: dead store & 1.2 & x & x & \checkmark & Infer \\
tube.c:86: error: dead store & 1.2 & x & x & \checkmark & Infer \\
job.c:95: error: dead store & 1.2 & x & x & \checkmark & Infer \\
conn.c:224: error: dead store & 1.2 & x & x & \checkmark & Infer \\
prot.c:1583: error: null dereference & 1.2 & x & x & x & Infer \\
cut.c:222: error: resource leak & 1.2 & 1.5 & x & x & Cppcheck \\
job:52:error: invalid dereference & 1.2 & x & x & \checkmark & Predator \\
binlog.c:176: error: buffer overflow & 1.4.1 & 1.4.2 & \checkmark & x & x \\
binlog.c:215: null dereference & 1.4.1 & 1.4.2 & x & x & Clang Static An. \\
binlog.c:723: error: null dereference & 1.4.1 & 1.5 & x & x & Infer \\
job.c:70: error: memory leak & 1.4.2 & 1.4.3 & \checkmark & x & Predator \\
prot.c:974: error: memory leak & 1.4.2 & 1.4.3 & \checkmark & x & x \\
prot.c:1017: error: memory leak & 1.4.3 & 1.4.4 & \checkmark & x & x \\
net.c:31: error: dead store & 1.5 & 1.9 & x & x & \thead{Clang Static An. \\ Infer} \\
heap.c:98: error: memory leak & 1.5 & x & x & \checkmark & Infer \\
walg.c:293: error: memory leak & 1.5 & x & x & \checkmark & Infer \\
walg.c:293: error: resource leak & 1.5 & x & x & \checkmark & Infer \\
walg.c:357: error: memory leak & 1.5 & x & x & \checkmark & Infer \\
walg.c:357: error: resource leak & 1.5 & x & x & \checkmark & Infer \\ 
walg.c:426: error: resource leak  & 1.5 & x & x & \checkmark & Infer \\
prot.c:514: error: null dereference & 1.5 & 1.8 & x & x & Infer \\
prot.c:554: error: null dereference & 1.5 & 1.8 & x & x & Infer \\
file.c:204: error: null dereference & 1.6 & 1.7 & \checkmark & x & Clang Static An. \\
file.c:325: error: null dereference & 1.6 & 1.7 & \checkmark & x & Clang Static An. \\
prot.c:501: error: null dereference & 1.6 & x & x & x & Infer \\ 
conn.c:244: error: invalid free & 1.8 & 1.9 & \checkmark & x & x \\
integ-test.c:343: error: invalid free & 1.8 & 1.9 & \checkmark & x & x\\ 
prot.c:710: error: invalid free & 1.8 & 1.9 & \checkmark & x & x \\
darwin.c:84: error: dead store & 1.8 & x & x & x & Infer \\ 
walg.c:416: dead store & 1.9 & 1.10 & \checkmark & x & \thead{Clang Static An. \\ Infer} \\
testheap.c:222: error: memory leak & 1.10 & x & x & x & \thead{Cppcheck \\ Predator} \\
	\bottomrule
\end{tabular}
\end{adjustbox}
\end{table}

\clearpage

\begin{table}[ht]
	\caption{Resultados da análise do software SDS.}
	\label{tab:hla:sds}
\centering
\begin{adjustbox}{width=1\textwidth}
\begin{tabular}{l c c c c c c}
	\toprule
	\textbf{Erro}   & 	\multicolumn{2}{c}{\textbf{Versão}} & \textbf{Reportado} & \textbf{\thead{Falso \\ positivo}} & \textbf{Ferramenta} \\
	& Identificado & Corrigido & & & \\
	\midrule
sds.c:303: error: memory leak & 1 & 2 & \checkmark & x & x \\
sds.c:470: error: memory leak & 1 & x & \checkmark & \checkmark & x \\
sds.c:882: error: memory leak & 1 & 2 & \checkmark & x & x \\
sds.c:881: error: memory leak & 1 & 2 & \checkmark & x & x \\
sds.c:159: error: memory leak & 1 & 2 & \checkmark & x & Cppcheck \\
sds.c:160: error: null dereference & 1 & 2 & \checkmark & x & Infer \\
sds.c:717: error: null dereference & 1 & x & \checkmark & x & x \\
sds.c:817: error: null dereference & 1 & x & x & x & Infer \\
sds.c:891: null pointer & 1 & 2 & x & x & Clang Static An. \\
sds.c: 894: null pointer & 1 & 2 & x & x & Clang Static An. \\
sds.c:63: error: invalid dereference & 1 & x & x & \checkmark & Predator\\
sds.c:147: error: invalid dereference & 1 & x & x & \checkmark & Predator\\
sds.c:793: error: memory leak & 1 & x & x & \checkmark & Predator\\
sds.c:92: error: null dereference & 2 & x & \checkmark & x & Infer \\
sds.c:1123: error: null dereference & 2 & x & x & x & Infer \\
sds.c:580: error: dead store & 2 & x & x & \checkmark & Infer \\
sds.c:136: error: invalid dereference & 2 & x & x & \checkmark & Predator \\
sds.c:98: error: dereferencing object of size 8B out of bounds & 2 & x & x & \checkmark & Predator \\
sds.c:96: error: dereferencing object of size 4B out of bounds & 2 & x & x & \checkmark & Predator \\
sds.c:94: error: dereferencing object of size 2B out of bounds & 2 & x & x & \checkmark & Predator \\
sds.c:92: error: dereferencing object of size 1B out of bounds & 2 & x & x & \checkmark & Predator \\
sds.c:161: error: invalid free & 2 & x & x & \checkmark & \thead{Clang Static An. \\ Predator} \\
sds.c:161: error: memory leak & 2 & x & x & \checkmark & Predator \\
sds.c:1121: error: memory leak & 2 & x & x & \checkmark & Predator \\
sds.h:87: error: dereference NULL value with offset -1B & 2 & x & x & \checkmark & Predator \\
sds.h:197 error: invalid dereference & 2 & x & x & \checkmark & Predator \\
sds.c:229: invalid free & 2 & x & x & \checkmark & Clang Static An. \\
sds.c:1240: dead store & 2 & x & x & x & \thead{Clang Static An. \\ Infer}  \\
	\bottomrule
\end{tabular}
\end{adjustbox}
\end{table}
\section{Padrões de erros}
\label{app:patterns}

\renewcommand{\arraystretch}{1.5}
\begin{table}[H]
	\caption{Padrões de erros.}
	\label{tab:hla:padroes}
\centering
\begin{tabularx}{1.0\textwidth}{X c c c c c}
	\toprule
	\textbf{Padrão} & \textbf{Erro} & \multicolumn{4}{c}{\textbf{Ferramenta}}\\ 
	& & Cppcheck & \thead{Clang Static \\ Analyzer} & Infer & Predator\\
	\midrule
Apontador inicializado com o valor \texttt{NULL} e desreferenciado na mesma função. & Desreferência nula & \checkmark & \checkmark & \checkmark & \checkmark \\
Apontador inicializado com o valor \texttt{NULL} noutra função e desreferenciado. & Desreferência nula & x & \checkmark & \checkmark & \checkmark \\
Falta de verificação após uma alocação de memória. & Desreferência nula & x & x & \checkmark & x \\
O valor passado como argumento à função de alocação de memória (\texttt{malloc} ou \texttt{calloc}) é do tipo \texttt{*ptr}. & \makecell[t]{Desreferência nula \\ Fuga de memória \\ Operação de \\ libertação inválida} & x & \checkmark & x & \checkmark \\
O valor escrito no endereço é inicializado com o valor \texttt{NULL} ou 0 e nunca é usado. & \makecell[t]{Valor do endereço \\ nunca é usado} & x & x & \checkmark & x \\
O valor escrito no endereço é inicializado com um valor diferente de \texttt{NULL} e 0 e nunca é usado. & \makecell[t]{Valor do endereço \\ nunca é usado} & x & \checkmark & \checkmark & x \\
O valor do apontador é alterado através da função \texttt{realloc}, sem garantir que esta tem sucesso (sem chamadas a funções). & Fuga de memória & \checkmark & x & \checkmark & x \\
O valor do apontador é alterado através da função \texttt{realloc}, sem garantir que esta tem sucesso (com chamadas a funções). & Fuga de memória & x & x & \checkmark & x \\
Alocação e libertação de memória explicita. & Fuga de memória & \checkmark & \checkmark & \checkmark & \checkmark \\
Alocação e libertação de memória não explicita. & Fuga de memória & x & \checkmark & \checkmark & \checkmark \\
Libertação da estrutura antes da libertação da memória alocada para apontadores que fazem parte da mesma. & Fuga de memória & x & \checkmark & \checkmark & \checkmark \\
\bottomrule
\end{tabularx}
\end{table}
\section{Exemplos mínimos}
\label{app:exemplos_minimos}

Este apêndice está organizado por tipo de erro, contendo cada secção listagens com o código dos exemplos mínimos e os relatórios das ferramentas utilizadas.

\subsection{Dead store}

\begin{figure}[htbp]
	\centering
	\caption{Relatório de erros devolvido pela ferramenta Clang Static Analyzer para o exemplo mínimo da Listagem~\ref{lst:dead_store_tp}.}
	\includegraphics[height=3in]{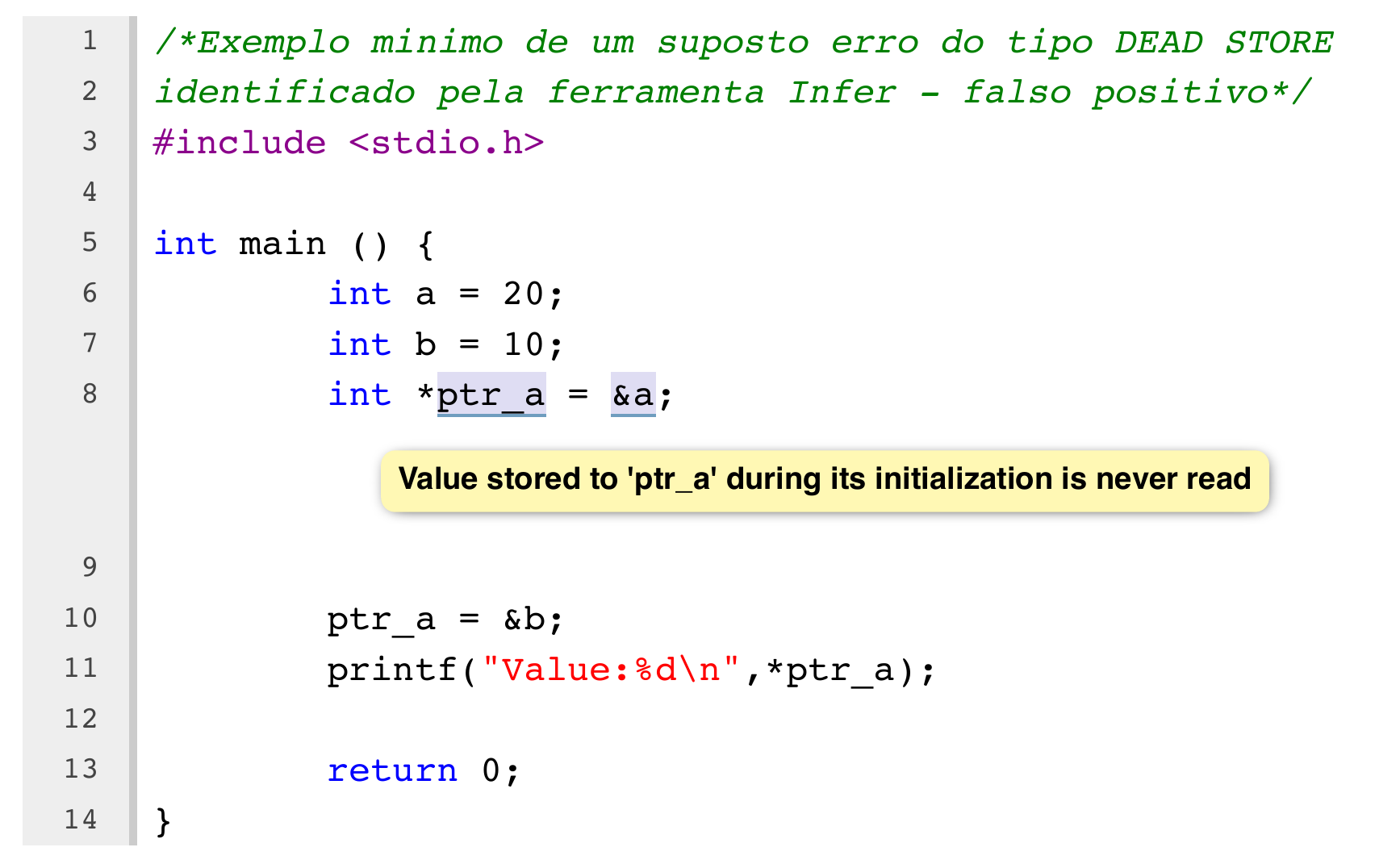}
	\label{fig:relatorio_dead_store_clang_static_analyzer}
\end{figure}

A Listagem~\ref{lst:dead_store_fp} corresponde ao exemplo mínimo de um falso positivo identificado pelo Infer (Listagem~\ref{lst:dead_store_fp_infer}). Este erro é considerado um falso positivo, porque o endereço é inicializado com o valor \texttt{NULL} e, portanto o valor do mesmo só é utilizado depois de lhe ser atribuído um novo valor. Caso o endereço tivesse sido inicializado com o valor 0, o Infer iria devolver o mesmo tipo de resultado. Quando este erro surge com este tipo de padrão não é reportado por mais nenhuma das restantes ferramentas. Na Listagem~\ref{lst:dead_store_corr} o erro é corrigido e o Infer não devolve qualquer tipo de resultado. Por fim, na Listagem~\ref{lst:dead_store_tp} está representado um exemplo de um erro do mesmo tipo, inicializado com um valor diferente de \texttt{NULL} ou 0, que é reportado por todas as ferramentas capazes de identificar erros do tipo \emph{dead store} (Listagem~\ref{lst:dead_store_tp_infer} e Figura~\ref{fig:relatorio_dead_store_clang_static_analyzer}). 

\lstset{language=C, caption=Exemplo mínimo de um erro 
    \emph{dead store} classificado como falso positivo., label=lst:dead_store_fp}
\begin{lstlisting}[frame=single] 
#include <stdio.h>

int main () {
	int var = 20;
	int *ptr_a = NULL;
	
	ptr_a = &var;
	printf("Value:%d\n",*ptr_a);

   	return 0;
}
\end{lstlisting}

\lstset{language=C, caption=Relatório de resultados d
0 Infer para o exemplo mínimo da Listagem~\ref{lst:dead_store_fp}., label=lst:dead_store_fp_infer}
\begin{lstlisting}[frame=single] 
Found 1 issue

dead_store_false_positive_infer.c:7: error: DEAD_STORE
  The value written to &ptr_a is never used.
  5.   int main () {
  6.   	int var = 20;
  7. > 	int *ptr_a = NULL;
  8.   	
  9.   	ptr_a = &var;

Summary of the reports

  DEAD_STORE: 1
\end{lstlisting}

\lstset{language=C, caption=Exemplo mínimo de um erro do tipo \emph{dead store} corrigido., label=lst:dead_store_corr}
\begin{lstlisting}[frame=single] 
#include <stdio.h>

int main () {
	int var = 20;
	int *ptr_a = &var;
	printf("Value:%d\n",*ptr_a);

   	return 0;
}

\end{lstlisting}

\lstset{language=C, caption=Exemplo mínimo de um erro do tipo \emph{dead store} classificado como verdadeiro positivo., label=lst:dead_store_tp}
\begin{lstlisting}[frame=single] 
#include <stdio.h>

int main () {
	int a = 20;
	int b = 10;
	int *ptr_a = &a;
	
	ptr_a = &b;
	printf("Value:%d\n",*ptr_a);

   	return 0;
}
\end{lstlisting}

\lstset{language=C, caption=Relatório de resultados da ferramenta Infer para o exemplo mínimo da Listagem~\ref{lst:dead_store_tp}., label=lst:dead_store_tp_infer}
\begin{lstlisting}[frame=single] 
Found 1 issue

dead_store_false_positive_infer.c:8: error: DEAD_STORE
  The value written to &ptr_a is never used.
  6.   	int a = 20;
  7.   	int b = 10;
  8. > 	int *ptr_a = &a;
  9.   	
  10.   	ptr_a = &b;

Summary of the reports

  DEAD_STORE: 1
\end{lstlisting}

\subsection{Fuga de memória}

\subsubsection{Argumento passado ao malloc.}
\begin{figure}[H]
	\centering
	\caption{Relatório de erros devolvido pela ferramenta Clang Static Analyzer para o exemplo mínimo da Listagem~\ref{lst:memory_leak_mallocarg_tp}.}
	\includegraphics[height=3in]{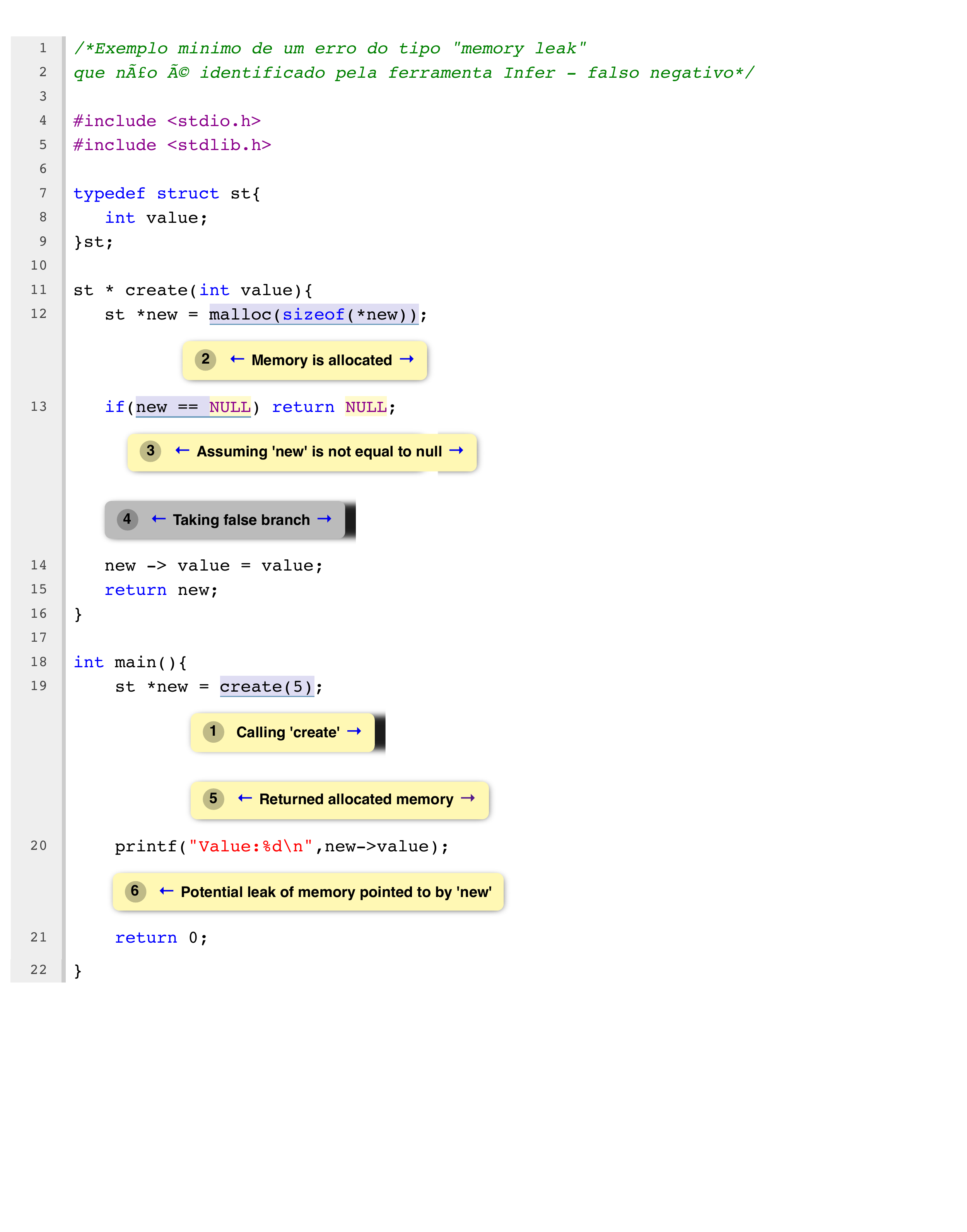}
	\label{fig:relatorio_memory_leak_mallocarg_clang_static_analyzer}
\end{figure}

\begin{figure}[H]
	\centering
	\caption{Relatório de erros devolvido pela ferramenta Clang Static Analyzer para o exemplo mínimo da Listagem~\ref{lst:memory_leak_mallocarg_tp_identificado}.}
	\includegraphics[height=3in]{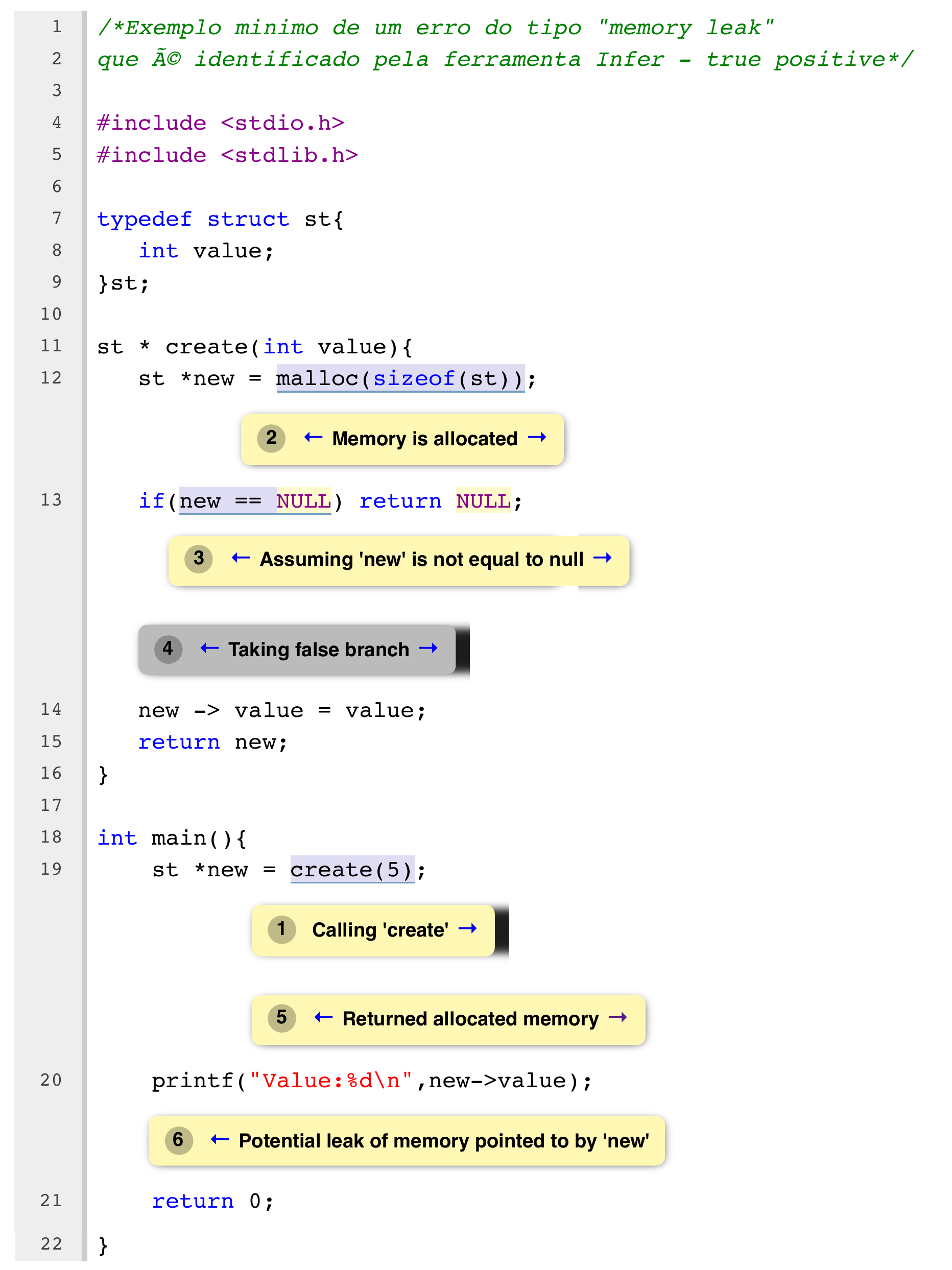}
	\label{fig:relatorio_memory_leak_mallocarg_identificado_clang_static_analyzer}
\end{figure}

Na Listagem~\ref{lst:memory_leak_mallocarg_tp} está representado o exemplo mínimo onde o argumento passado à função \texttt{malloc} é do tipo \texttt{ptr*} e, portanto, o Infer (até à versão 0.13.1) não é capaz de identificar a fuga de memória presente no código.  Por outro, o Predator e o Clang Static Analyzer conseguem identificar o erro, como pode ser observado no relatório de resultados da Listagem~\ref{lst:memory_leak_mallocarg_tp_results_predator} e da Figura~\ref{fig:relatorio_memory_leak_mallocarg_clang_static_analyzer}.
Na Listagem~\ref{lst:memory_leak_mallocarg_tp_corr} a fuga de memória foi corrigida e não é reportada por nenhuma das ferramentas. 
De notar que, qualquer erro que exista no código depois da chamada à função \texttt{malloc}, passando o argumento \texttt{ptr*}, é ignorado pelo Infer. 
Na Listagem~\ref{lst:memory_leak_mallocarg_tp_identificado} o argumento passado à função \texttt{malloc} é do mesmo tipo da estrutura utilizada, logo, qualquer versão do Infer é capaz de identificar a fuga de memória presente no código (Listagem~\ref{lst:memory_leak_mallocarg_tp_results_infer}). Além disso, o Predator e o Clang Static Analyzer (Listagem~\ref{lst:relatorio_memory_leak_mallocarg_identificado_predator} e Figura~\ref{fig:relatorio_memory_leak_mallocarg_identificado_clang_static_analyzer}) também conseguem identificar o erro, tal como já acontecia no exemplo mínimo anterior. 
\newpage
\lstset{language=C, caption=Exemplo mínimo de um erro do tipo fuga de memória., label=lst:memory_leak_mallocarg_tp}
\begin{lstlisting}[frame=single] 
#include <stdio.h>
#include <stdlib.h>

typedef struct st{
   int value;
}st;

st * create(int value){
   st *new = malloc(sizeof(*new));
   if(new == NULL) return NULL;
   new -> value = value;
   return new;
}

int main(){	
    st *new = create(5);
    printf("Value:%d\n",new->value);
    return 0;
}
\end{lstlisting}

\lstset{language=C, caption=Relatório de resultados da ferramenta Predator para o exemplo mínimo da Listagem~\ref{lst:memory_leak_mallocarg_tp}., label=lst:memory_leak_mallocarg_tp_results_predator}
\begin{lstlisting}[frame=single] 
memory_leak_false_negative_infer.c:20:11: warning: memory 
leak detected while destroying a variable on stack 
[-fplugin=libsl.so]
/home/fct/predator/cl/cl_easy.cc:83: note: clEasyRun() 
took 0.001 s [internal location] [-fplugin=libsl.so]
\end{lstlisting}

\lstset{language=C, caption=Exemplo mínimo de um erro do tipo fuga de memória corrigido., label=lst:memory_leak_mallocarg_tp_corr}
\begin{lstlisting}[frame=single] 
#include <stdio.h>
#include <stdlib.h>

typedef struct st{
   int value;
}st;

st * create(int value){
   st *new = malloc(sizeof(st));
   if(new == NULL) return NULL;
   new -> value = value;
   return new;
}

int main(){	
    st *new = create(5);
    free(new);
    return 0;
}
\end{lstlisting}

\lstset{language=C, caption=Exemplo mínimo de um erro do tipo fuga de memória identificado pelo Infer., label=lst:memory_leak_mallocarg_tp_identificado}
\begin{lstlisting}[frame=single] 
#include <stdio.h>
#include <stdlib.h>

typedef struct st{
   int value;
}st;

st * create(int value){
   st *new = malloc(sizeof(st));
   if(new == NULL) return NULL;
   new -> value = value;
   return new;
}

int main(){	
    st *new = create(5);
	if(new){
		printf("Value:%d\n",new->value);
	}
    return 0;
}
\end{lstlisting}

\lstset{language=C, caption=Relatório de resultados da ferramenta Infer para o exemplo mínimo da Listagem~\ref{lst:memory_leak_mallocarg_tp_identificado}., label=lst:memory_leak_mallocarg_tp_results_infer}
\begin{lstlisting}[frame=single] 
Found 1 issue

memory_leak_true_positive_infer.c:21: error: MEMORY_LEAK
  memory dynamically allocated to `new` by call to 
  `create()` at line 19, column 15 is not reachable 
  after line 21, column 5.
  19.       st *new = create(5);
  20.       if(new){
  21. >     printf("Value:%d\n",new->value);
  22.       }
  23.       return 0;

Summary of the reports

  MEMORY_LEAK: 1
\end{lstlisting}
\newpage
\lstset{language=C, caption=Relatório de resultados da ferramenta Predator para o exemplo mínimo da Listagem~\ref{lst:memory_leak_mallocarg_tp_identificado}., label=lst:relatorio_memory_leak_mallocarg_identificado_predator}
\begin{lstlisting}[frame=single] 
memory_leak_true_positive_infer.c:20:11: warning: memory 
leak detected while destroying a variable on stack 
[-fplugin=libsl.so]
/home/fct/predator/cl/cl_easy.cc:83: note: clEasyRun() 
took 0.001 s [internal location] [-fplugin=libsl.so]
\end{lstlisting}

\subsubsection{Estrutura com apontadores.}

Na Listagem~\ref{lst:memory_leak_structwithpointers} está representado o exemplo mínimo de um erro do tipo fuga de memória que é identificado por todas as ferramentas (consultar a Listagem~\ref{lst:memory_leak_structwithpointers_results_cppcheck}, a Listagem~\ref{lst:memory_leak_structwithpointers_results_infer} e a Listagem~\ref{lst:memory_leak_structwithpointers_results_predator}), exceto pelo Clang Static Analyzer. Na Listagem~\ref{lst:memory_leak_structwithpointers_corr} a fuga de memória foi corrigida e, portanto, não é reportada por nenhuma das ferramentas.
\newpage
\lstset{language=C, caption=Exemplo mínimo de um erro do tipo fuga de memória numa estrutura com apontadores., label=lst:memory_leak_structwithpointers}
\begin{lstlisting}[frame=single] 
#include <stdio.h>
#include <stdlib.h>

typedef struct st{
   int *value;
}st;

int main(){	
    st *new = malloc(sizeof(st));
	if(!new) return -1;
	new->value = malloc(sizeof(int));
	free(new);
	return 0;
}
\end{lstlisting}

\lstset{language=C, caption=Relatório de erros devolvido pela ferramenta Cppcheck para o exemplo mínimo da Listagem~\ref{lst:memory_leak_structwithpointers}., label=lst:memory_leak_structwithpointers_results_cppcheck}
\begin{lstlisting}[frame=single] 
Checking memory_leak_true_positive_structwithpointer.c ...
[memory_leak_true_positive_structwithpointer_infer.c:19]: 
(error) Memory leak: new.value
\end{lstlisting}

\lstset{language=C, caption=Relatório de erros devolvido pela ferramenta Infer para o exemplo mínimo da Listagem~\ref{lst:memory_leak_structwithpointers}., label=lst:memory_leak_structwithpointers_results_infer}
\begin{lstlisting}[frame=single] 
Found 1 issue

memory_leak_true_positive_structwithpointer.c:19: 
error: MEMORY_LEAK memory dynamically allocated to 
`new->value` by call to `malloc()` at line 14, 
column 15 is not reachable after line 19, column 2.
  17.   		return -1;
  18.   	}*/
  19. > 	free(new);
  20.   	return 0;

Summary of the reports

  MEMORY_LEAK: 1
\end{lstlisting}
\newpage
\lstset{language=C, caption=Relatório de erros devolvido pela ferramenta Predator para o exemplo mínimo da Listagem~\ref{lst:memory_leak_structwithpointers}.,
label=lst:memory_leak_structwithpointers_results_predator}
\begin{lstlisting}[frame=single] 
memory_leak_true_positive_structwithpointer.c:19:6: 
warning: memory leak detected while destroying a 
heap object [-fplugin=libsl.so]
/home/fct/predator/cl/cl_easy.cc:83: note: clEasyRun() 
took 0.001 s [internal location] [-fplugin=libsl.so]
\end{lstlisting}

\lstset{language=C, caption=Exemplo mínimo de um erro do tipo fuga de memória numa estrutura com apontadores corrigido., label=lst:memory_leak_structwithpointers_corr}
\begin{lstlisting}[frame=single] 
#include <stdio.h>
#include <stdlib.h>

typedef struct st{
   int *value;
}st;

int main(){	
    st *new = malloc(sizeof(st));
	if(!new) return -1;
	new->value = malloc(sizeof(int));
	free(new->value);
	free(new);
	return 0;
}
\end{lstlisting}

\subsubsection{Falta de verificação do realloc}
A falta de verificação de uma operação de memória é um tipo de erro apenas identificado pelo Infer. No entanto, quando se trata da verificação de uma chamada à função \texttt{realloc} (Listagem~\ref{lst:memory_leak_realloc}) o Cppcheck identifica um erro do tipo fuga de memória, no caso do exemplo mínimo apresentado nesta secção o Infer também identifica um erro deste tipo. Os relatórios de resultados do Infer e do Cppcheck encontram-se na Listagem~\ref{lst:memory_leak_realloc_results_cppcheck} e na Listagem~\ref{lst:memory_leak_realloc_results_infer}, respetivamente. O Clang Static Analyzer não reporta qualquer resultado para este exemplo. O Predator devolve um falso positivo, uma vez que esta ferramenta ignora a chamada à função \texttt{realloc}.
Na Listagem~\ref{lst:memory_leak_realloc_corr} encontra-se representada a correção da fuga de memória do exemplo anterior, sendo que não é reportado qualquer tipo de erro pelas ferramentas, com a exceção do Predator que, como já foi referido, reporta um falso positivo.
\newpage
\lstset{language=C, caption=Exemplo mínimo de um erro do tipo fuga de memória utilizando uma função \texttt{realloc}., label=lst:memory_leak_realloc}
\begin{lstlisting}[frame=single] 
#include <stdio.h>
#include <stdlib.h>

typedef struct st{
   int value;
}st;

int main(){	
    st *new = malloc(sizeof(st));
	if(!new) return -1;
	new = realloc(new,2*sizeof(st));
	free(new);
	return 0;
}
\end{lstlisting}

\lstset{language=C, caption=Relatório de erros devolvido pela ferramenta Cppcheck para o exemplo mínimo da Listagem~\ref{lst:memory_leak_realloc}., label=lst:memory_leak_realloc_results_cppcheck}
\begin{lstlisting}[frame=single] 
Checking memory_leak_true_positive_realloc.c ...
[memory_leak_true_positive_realloc_infer.c:14]: 
(error) Common realloc mistake: 'new' nulled 
but not freed upon failure
\end{lstlisting}

\lstset{language=C, caption=Relatório de erros devolvido pela ferramenta Infer para o exemplo mínimo da Listagem~\ref{lst:memory_leak_realloc}., label=lst:memory_leak_realloc_results_infer}
\begin{lstlisting}[frame=single] 
Found 1 issue

memory_leak_true_positive_realloc_infer.c:14: error: 
MEMORY_LEAK memory dynamically allocated by call to 
`malloc()` at line 12, column 15 is not reachable 
after line 14, column 2.
  12.       st *new = malloc(sizeof(st));
  13.   	if(!new) return -1;
  14. > 	new = realloc(new,2*sizeof(st));
  15.   	free(new);
  16.   	return 0;

Summary of the reports

  MEMORY_LEAK: 1
\end{lstlisting}
\newpage
\lstset{language=C, caption=Exemplo mínimo de um erro do tipo fuga de memória utilizando uma função \texttt{realloc} corrigido., label=lst:memory_leak_realloc_corr}
\begin{lstlisting}[frame=single] 
#include <stdio.h>
#include <stdlib.h>

typedef struct st{
   int value;
}st;

int main(){	
	st *tmp;
    st *new = malloc(sizeof(st));
	if(!new) return -1;
	tmp = realloc(new,2*sizeof(st));
	if(tmp)
	{
		new = tmp;
	}
	free(new);
	return 0;
}
\end{lstlisting}

\subsection{Desreferência nula}
O Infer é a única ferramenta que reporta um erro quando não é feita a verificação da chamada a uma função de alocação de memória. Portanto, no exemplo mínimo da Listagem~\ref{lst:null_dereference}, apenas foram obtidos relatórios de resultados do Infer (Listagem~\ref{lst:null_dereference_infer}). Na Listagem~\ref{lst:null_dereference_corr} o erro é corrigido e não deixa de ser reportado pelo Infer.
\newpage
\lstset{language=C, caption=Exemplo mínimo de um erro do tipo desreferência nula., label=lst:null_dereference}
\begin{lstlisting}[frame=single] 
#include <stdio.h>
#include <stdlib.h>

typedef struct st{
   int value;
}st;

st * create(int value){
   st *new = malloc(sizeof(st));
   new -> value = value;
   return new;
}

int main(){	
    st *new = create(5);
    free(new);
    return 0;
}
\end{lstlisting}

\lstset{language=C, caption=Relatório de erros devolvido pela ferramenta Infer para o exemplo mínimo da Listagem~\ref{lst:null_dereference}., label=lst:null_dereference_infer}
\begin{lstlisting}[frame=single] 
Found 1 issue

null_dereference_true_positive_mallocverification.c:13: 
error: NULL_DEREFERENCE pointer `new` last assigned on 
line 12 could be null and is dereferenced at line 13, 
column 4.
  11.   st * create(int value){
  12.      st *new = malloc(sizeof(st));
  13. >    new -> value = value;
  14.      return new;
  15.   }

Summary of the reports

  NULL_DEREFERENCE: 1
\end{lstlisting}
\newpage
\lstset{language=C, caption=Exemplo mínimo de um erro do tipo desreferência nula corrigido., label=lst:null_dereference_corr}
\begin{lstlisting}[frame=single] 
#include <stdio.h>
#include <stdlib.h>

typedef struct st{
   int value;
}st;

st * create(int value){
   st *new = malloc(sizeof(st));
   if(!new) return NULL;
   new -> value = value;
   return new;
}

int main(){	
    st *new = create(5);
    free(new);
    return 0;
}
\end{lstlisting}

\section{Síntese dos relatórios de resultados das ferramentas}
\label{app:sintese_erros}

\begin{table}[ht]
	\caption{Erros identificados no SDS.}
	\label{tab:hla:erros-sds}
\centering
\begin{adjustbox}{width=1\textwidth}
\begin{tabular}{l c c c c c c c c c c c}
	\toprule
	\textbf{Tipo de erro} &
	\multicolumn{2}{c}{\textbf{Cppcheck}} &
	\multicolumn{2}{c}{\textbf{\thead{Clang Static An.}}} &
	\multicolumn{2}{c}{\textbf{Infer}} &
	\multicolumn{2}{c}{\textbf{Predator}} &
	\textbf{\thead{Não identificados \\ p/ ferramentas}} &
	\textbf{\thead{Identificados \\ apenas p/ uma \\ ferramenta}} &
	\textbf{Total}\\
	& FP & TP & FP & TP & FP & TP & FP & TP\\
	\midrule
Memory leak & 0 & 1 & 0 & 0 & 0 & 0 & 3 & 0 & 4 & 1(Cppcheck) & 8\\
Null dereference & 0 & 0 & 0 & 2 & 0 & 4 & 1 & 0 & 1 & 4(Infer) + 2(CSA) & 8\\
Invalid dereference & 0 & 0 & 0 & 0 & 0 & 0 & 4 & 0 & 1 & 0 & 4\\
Invalid free & 0 & 0 & 2 & 0 & 0 & 0 & 1 & 0 & 0 & 0 & 2\\
Dead store & 0 & 0 & 1 & 0 & 1 & 1 & 0 & 0 & 0 & 0 & 2\\
Out of bounds & 0 & 0 & 0 & 0 & 0 & 0 & 4 & 0 & 0 & 0 & 4\\
	\bottomrule
\end{tabular}
\end{adjustbox}
\end{table}

\begin{table}[ht]
	\caption{Erros identificados no Beanstalkd.}
	\label{tab:hla:erros-beanstalkd}
\centering
\begin{adjustbox}{width=1\textwidth}
\begin{tabular}{l c c c c c c c c c c c}
	\toprule
	\textbf{Tipo de erro} &
	\multicolumn{2}{c}{\textbf{Cppcheck}} &
	\multicolumn{2}{c}{\textbf{\thead{Clang Static An.}}} &
	\multicolumn{2}{c}{\textbf{Infer}} &
	\multicolumn{2}{c}{\textbf{Predator}} &
	\textbf{\thead{Não identificados \\ p/ ferramentas}} &
	\textbf{\thead{Identificados \\ apenas p/ uma \\ ferramenta}} &
	\textbf{Total}\\
	& FP & TP & FP & TP & FP & TP & FP & TP\\
	\midrule
Memory leak & 0 & 1 & 0 & 0 & 4 & 1 & 3 & 2 & 7 & 1(Infer) + 1(Predator) & 17\\
Resource leak & 0 & 1 & 0 & 0 & 2 & 2 & 0 & 0 & 0 & 2(Infer) + 1(Cppcheck) & 5\\
Null dereference & 0 & 0 & 0 & 3 & 0 & 9 & 1 & 0 & 0 & 9(Infer) + 3(CSA) & 13\\
Invalid dereference & 0 & 0 & 0 & 0 & 0 & 0 & 125 & 0 & 0 & 0 & 125\\
Invalid free & 0 & 0 & 0 & 0 & 0 & 0 & 1 & 0 & 3 & 0 & 4\\
Dead store & 0 & 0 & 0 & 3 & 8 & 4 & 0 & 0 & 0 & 1(Infer) & 12\\
Buffer overflow & 0 & 0 & 0 & 0 & 0 & 0 & 0 & 0 & 1 & 0 & 1\\
Dangling pointer & 0 & 0 & 0 & 0 & 0 & 0 & 0 & 0 & 1 & 0 & 1\\
	\bottomrule
\end{tabular}
\end{adjustbox}
\end{table}
\section{Tempos de execução das ferramentas}
\label{app:tempos_execucao}

\begin{table}[ht]
	\caption{Tempos de execução em segundos 
	das ferramentas na análise do SDS.}
	\label{tab:hla:tempos_execucao_sds}
\centering
\begin{tabular}{l c c c c }
	\toprule
	\textbf{Versão} & 
	\textbf{Cppcheck} & 
	\textbf{\thead{Clang}} &
	\textbf{Infer} &
	\textbf{Predator} \\
	\midrule
1.0.0 & 0,292 & 5,961 & 12,562 & 0,409\\
2.0.0 & 0,75 & 9,037 & 14,891 & 1,075 \\
	\bottomrule
\end{tabular}
\end{table}

\begin{figure}
    \centering
   \includegraphics[width=\linewidth]{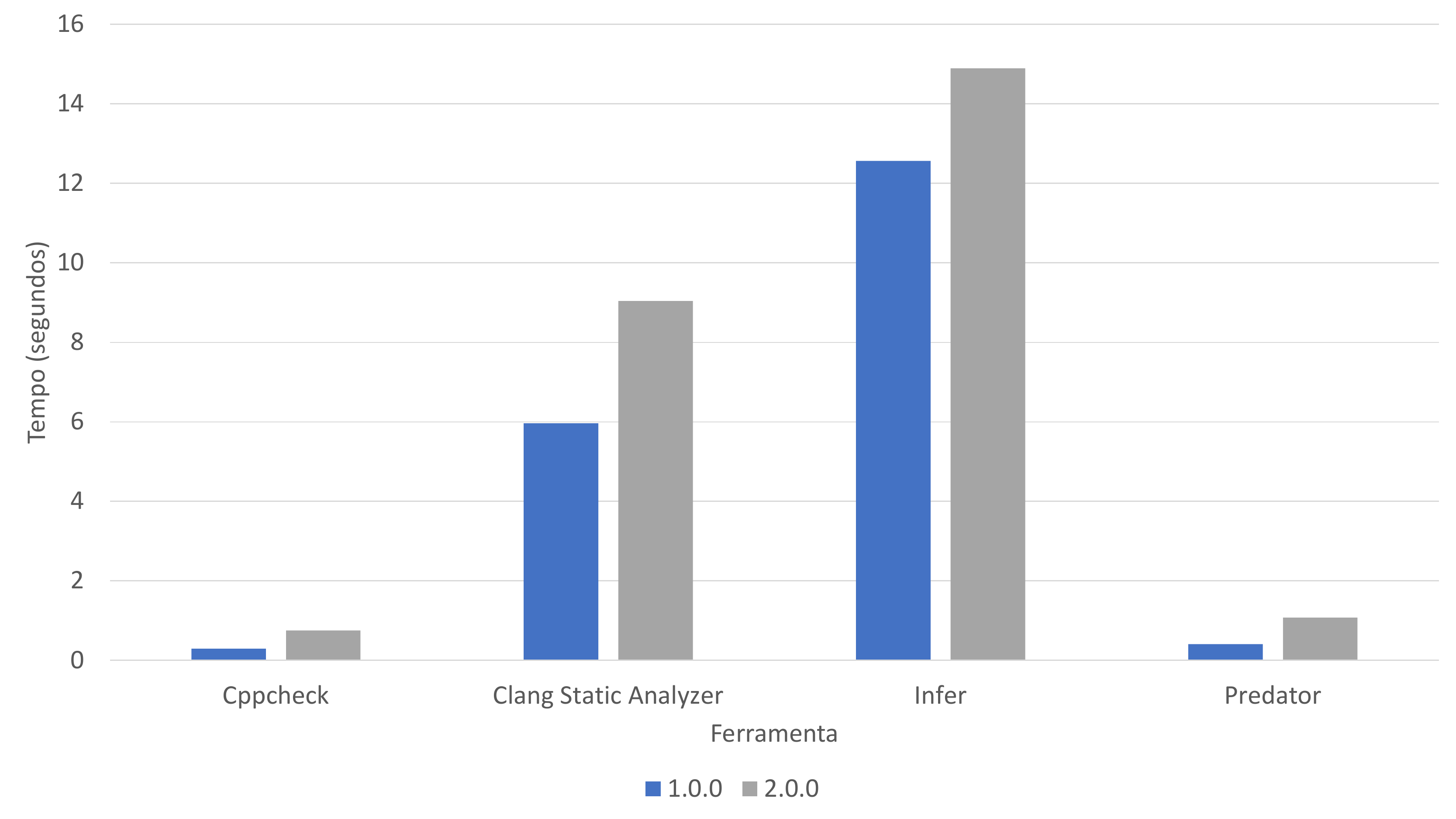}
   \caption{Tempos de execução em segundos 
   das ferramentas na análise do SDS.}
   \label{fig:tempos_execucao_sds}
\end{figure}

\begin{table}
	\caption{Tempos de execução em segundos 
	das ferramentas na análise do Beanstalkd.}
	\label{tab:hla:tempos_execucao_beanstalkd}
\centering
\begin{tabular}{l c c c c }
	\toprule
	\textbf{Versão} & 
	\textbf{Cppcheck} & 
	\textbf{\thead{Clang}} &
	\textbf{Infer} &
	\textbf{Predator} \\
	\midrule
0.3 & 0,397 & 3,489 & 40,881 & 0,673 \\
0.4 & 0,401 & 3,311 & 42,349 & 0,682 \\
0.8 & 0,657 & 2,683 & 51,502 & 0,81 \\
1.2 & 1,075 & 11,453 & 59,923 & 1,06 \\
1.4.1 & 1,152 & 13,027 & 31,358 & 0,727 \\
1.4.2 & 1,14 & 12,685 & 37,149 & 0,761 \\
1.4.3 & 1,179 &	13,631 & 32,641	& 0,773 \\
1.5	& 1,831	& 5,323	& 26,556 &	2,379 \\
1.6	& 2,241	& 15,923 & 63,914 &	2,421 \\
1.8	& 2,144	& 5,194	& 27,499 & 2 \\
1.9	& 2,031	& 15,93	& 71,174 & 2,116 \\
1.10 & 2,255 & 15,982 & 64,306 & 2,365 \\
	\bottomrule
\end{tabular}
\end{table}

\begin{figure}
   \centering
   \includegraphics[width=\linewidth]{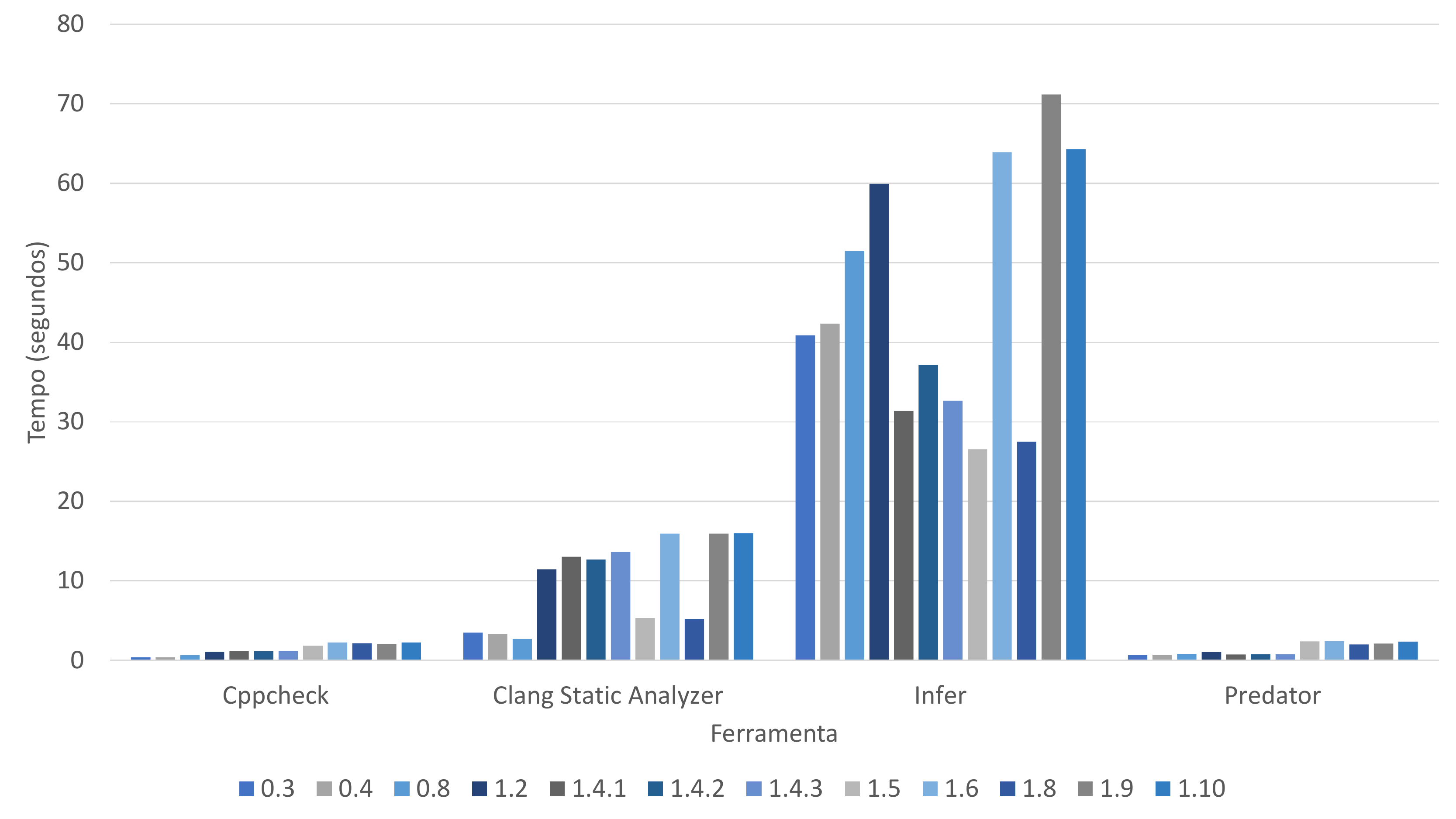}
   \caption{Tempos de execução em segundos 
   das ferramentas na análise do Beanstalkd.}
   \label{fig:tempos_execucao_beanstalkd}
\end{figure}
\newpage
\section{Tempo de permanência dos erros nos softwares.}
\label{app:tempos_permanencia}

\begin{table}[ht]
	\caption{Datas e tempo em meses decorrido desde a introdução até à correção dos erros no SDS.}
	\label{tab:hla:tempos_permanencia_sds}
\centering
\begin{adjustbox}{width=1\textwidth}
\begin{tabular}{l c c c}
	\toprule
	\textbf{Erro} & 
	\textbf{Introduzido} & 
	\textbf{Corrigido} &
	\textbf{Intervalo de tempo} \\ 
	\midrule
sds.c:159: error: memory leak & 06/02/2014 & 25/11/2014 & 9 \\
sds.c:160: error: null dereference & 06/02/2014	& 25/11/2014 & 9 \\
sds.c:817: error: null dereference & 06/02/2014	& x & x \\
sds.c:891: null dereference	& 06/02/2014 & 25/07/2015 & 17 \\
sds.c: 894: null dereference & 06/02/2014 & 25/07/2015 & 17 \\
sds.c:92: error: null dereference & 25/07/2015 & 07/02/2018 & 30 \\
sds.c:1123: error: null dereference	& 25/07/2015 & x & x \\
sds.c:1240: dead store	& 25/07/2015	& x & x \\
	\bottomrule
\end{tabular}
\end{adjustbox}
\end{table}

\begin{table}[ht]
	\caption{Datas e tempo em meses decorrido desde a introdução até à correção dos erros no Beanstalkd.}
	\label{tab:hla:tempos_permanencia_beanstalkd}
\centering
\begin{adjustbox}{width=1\textwidth}
\begin{tabular}{l c c c }
	\toprule
	\textbf{Erro} & 
	\textbf{Introduzido} & 
	\textbf{Corrigido} &
	\textbf{Intervalo de tempo} \\ 
	\midrule
net.c:28: error: dead store & 08/11/2007 & 03/10/2009 & 22 \\
beanstalkd.c:41: error: resource leak &	08/11/2007 &	28/01/2012 &	50 \\
reserve.c:51: error: null dereference & 08/11/2007 & 28/01/2012 & 50 \\
prot.c:140: error:  null dereference & 08/11/2007 & 28/01/2012 & 50 \\
beanstalkd.c:395: error: null dereference &	08/11/2007 & 17/01/2009 & 14 \\
beanstalkd.c:737: error: memory leak & 08/11/2007 & 11/12/2007 & 1 \\
prot.c:320: error: null dereference & 01/02/2008 & 28/01/2012 & 47 \\
prot.c:374: error: null dereference & 01/02/2008	& 28/01/2012	& 47 \\
prot.c:1583: error: null dereference & 17/01/2009 & x & x \\
cut.c:222: error: resource leak & 17/02/2009 & 28/01/2012 & 36 \\
binlog.c:215: null dereference & 14/10/2009 & 18/10/2009 & 0,13 \\
binlog.c:723: error: null dereference & 14/10/2009 & 28/01/2012 & 27 \\
job.c:70: error: memory leak & 18/10/2009 & 30/11/2009 & 1 \\
net.c:31: error: dead store & 28/01/2012 & 14/04/2013 & 14 \\
prot.c:514: error: null dereference & 28/01/2012 & 03/11/2012 & 9 \\
prot.c:554: error: null dereference & 28/01/2012 & 03/11/2012 & 9 \\
walg.c:426: error: resource leak & 28/01/2012 & x & x \\
file.c:204: error: null dereference & 10/05/2012 & 02/09/2012 & 3 \\
file.c:325: error: null dereference & 10/05/2012 & 02/09/2012 & 3 \\
prot.c:501: error: null dereference & 10/05/2012 & x & x \\
darwin.c:84: error: dead store & 03/11/2012 & x & x \\
walg.c:416: dead store & 13/04/2013 & 05/08/2014 & 15 \\
testheap.c:222: error: memory leak & 05/08/2014 & x & x \\
	\bottomrule
\end{tabular}
\end{adjustbox}
\end{table}

\end{document}